\documentclass{jpp}
\usepackage{graphicx}
\usepackage{placeins}
\usepackage[utf8]{inputenc}
\usepackage[T1]{fontenc}
\usepackage{amsmath}

\shorttitle{Homo-interaction between a pair of rising/falling bubbles/droplets}
\shortauthor{Dharodi $et~al.$}

\title{A numerical study of gravity-driven instability in strongly coupled dusty plasma. Part 3: Homo-interaction between a pair of rising/falling bubbles/droplets}

 \author{Vikram S. Dharodi \aff{1}
 \corresp{\email{vsd0005@auburn.edu}},\aff{2}
  \and Evdokiya Kostadinova \aff{1}}

\affiliation{\aff{1}Department of Physics, Auburn University, Auburn, Alabama 32849, USA \aff{2} Department of Physics and Astronomy, West Virginia University, Morgantown, WV, USA\corresp{Current affiliation}}

\begin{document}

\maketitle

\begin{abstract}
A numerical study of the homo-interactions between two falling droplets and between two rising bubbles in a strongly coupled dusty plasma medium is presented in this article. This strongly coupled dusty plasma is considered as a viscoelastic fluid using the generalized hydrodynamic fluid model formalism. Two factors that affect homo-interactions are taken into account: the initial spacing and the coupling strength of the medium. Three different spacings between two droplets are simulated: widely, medium, and closely. In each case, the coupling strength has been given as mild-strong and strong. It is shown that the overall dynamic is governed by the competition between the acceleration of two droplets/bubbles due to gravity and the interaction due to the closeness of the droplets/bubbles. Especially in viscoelastic fluids, the closeness between two droplets/bubbles, aside from their initial separation, at a later time may result from shear waves that emerge from rotating vorticies. For widely-spaced, unlike classical hydrodynamic fluids, we find that shear waves in viscoelastic fluids facilitate the pairing between two bubbles/droplets. In the case of medium-spaced, the two new dipoles of unequal strength blobs exhibit a circular motion and exchange their partners. For closely-spaced, the droplet/bubble fall/rise is suppressed as the medium's coupling strength increases. Numerous two-dimensional simulations have been carried out. This work is a continuation of the work done in parts I (V. S. Dharodi and A. Das, J. Plasma Phys. 87 (02), 905870216 (2021)) and II (V. S. Dharodi, J. Plasma Phys. 87 (04), 905870402 (2021)).
\end{abstract}

\section{Introduction}\label{Introduction}

The buoyancy instability, which is driven by gravity, is what makes an object sinks or floats in a fluid. A droplet sinks into the fluid in the direction of gravity, whereas a bubble defies gravity and rises to the surface. Bubbles and droplets are frequently observed in a variety of fields, such as engineering , biomedical, agricultural, industrial, and environmental~\citep{moghtadernejad2020introduction,aliabouzar2023bubble,jia2015dynamics,wang2023sustained,zhao2017droplet,bourouiba2012drops}. From fundamental perspective of physics, droplets and bubbles play a significant role in transportation across any medium, for example, heat exchange, diffusion, mixing, etc. Understanding the interaction between bubbles and droplets is therefore essential to comprehending how the flow field evolves. Bubbles and droplets have been studied separately and together \citep{shew2006viscoelastic,gaudron2015bubble,dollet2019bubble,dwyer1989calculations,cristini2004theory,zhu2008three,leong2020droplet,mokhtarzadeh1985dynamics,chen2008droplet,tabor2011homo}. Numerical simulations and experiments have been carried out recently to study the interactions between two bubbles rising side by side in viscous liquids~\citep{zhang2019vortex,kong2019hydrodynamic}. Using atomic force microscopy, the homo- and hetero-interactions between oil droplets and air bubbles are assessed by~\cite{tabor2011homo}. In order to study the plasma-liquid interaction, water droplet evolution in a plasma medium and underwater discharge in the bubbles have been studied by \cite{oinuma2020controlled} and \cite{tachibana2011analysis}, respectively. In plasmas, studies of bubbles and/or droplets under different circumstances have been carried out \citep{arzhannikov2013surface,wang2015ionospheric,ning2021propagation,ou2021bubble}.~\cite{stenzel2012oscillatinga,stenzel2012oscillatingb,stenzel2012oscillatingc,stenzel2012oscillatingd} conducted a series of experiments on oscillating plasma bubbles, where bubbles are produced by introducing a negatively biased grid into the surrounding plasma. A laboratory dusty plasma is a good illustration of a homogeneous plasma; however, density fluctuations may be created using laser pulses and external forces like magnetic and/or electric, size or charge imbalance, etc. These density fluctuations can be associated with bubble and/or droplet conditions. The dynamics of bubbles have been investigated experimentally by \cite{chu2003observation} and \cite{teng2008micro}. \cite{chen2006interaction} studied the bubble-bubble interactions. Under the effect of thermophoresis, \cite{schwabe2009formation} has reported the spontaneous creation of bubbles, drops, and spraying cusps. 

In order to create a bubble and a droplet, we added a low-density region and a high-density region to the background fluid, respectively. In Part I~\citep{dharodi2021numericalA} of this paper, we separately explored the dynamics of a rising bubble and a falling droplet, and in Part II \citep{dharodi2021numericalB}, we explored the hetero- (bubble-droplet) interactions between a rising bubble and a falling droplet. For these studies, we have considered the dusty plasma as a medium. The study of dust is important from the perspective of both industrial \citep{melzer2000laser,merlino2004dusty,chaubey2023controlling,chaubey2023mitigating,ramkorun2023introducing,ramkorun2024comparing} and fundamental physics \citep{choudhary2016propagation,choudhary2017experimental,choudhary2018collective, dharodi2023ring}. Dusty plasma is gravitationally favorable due to the high mass of dust grains and readily forms a strong coupling regime caused by the high charge of dust grains. Therefore, the strongly coupled dusty plasma (SCDP) can serve as a perfect medium to study the role of strong coupling on the growth of gravity-driven instability like buoyancy. A SCDP behaves similarly to a VE fluid below the crystallization limit \citep{ikezi1986coulomb,vladimirov1997vibrational}. Here, we simulate the SCDP as VE fluid using a well-known phenomenological generalized hydrodynamics (GHD) fluid model \citep{Kaw_Sen_1998, Kaw_2001}. This model characterizes the viscoelastic effects through two coupling parameters: shear viscosity $\eta$ and the Maxwell relaxation parameter $\tau_m$ \citep{frenkel_kinetic}. Due to the strong correlation between dust grains, the viscosity $\eta$ in the presence of elasticity $\tau_m$ contributes to the transverse mode. This mode travels with a velocity $\sqrt{{\eta}/{\tau_m}}$. We study the incompressible limit of the GHD model (i-GHD) to investigate the exclusive effect of transverse modes and to avoid the possible coupling with the longitudinal mode. The transverse modes in the dusty plasma medium has been investigated by computational \citep{schmidt1997longitudinal}, experimental \citep{nunomura2000transverse, pramanik2002experimental}, and analytical \citep{peeters1987wigner,vladimirov1997vibrational,wang2001longitudinal} methods. Using the i-GHD model, we have computationally observed the shear waves and their role in various physical processes~\citep{dharodi2014visco,das2014collective,tiwari2014evolution,tiwari2014kelvin,tiwari2015turbulence,dharodi2016sub,dharodi2020rotating, dharodi2022kelvin,dharodi2024vortex,dharodicollective}. Employing the same model, in the current work, we examine how the coupling strength of the background VE fluid affects the homo-interactions between two falling droplets and between two rising bubbles. The idea of homo-interaction was proposed in Part II.

Three different spacings between two droplets are simulated: widely, medium, and closely. In each case, the coupling strength has been presented as mild-strong ($\eta$=2.5, $\tau_m$=20) and strong ($\eta$=2.5, $\tau_m$=10). Since we observe that the behavior of two falling droplets is the similar 
as the behavior of two rising bubbles so will discuss here mainly droplet case. The emerging shear waves from the rotating vorticity lobes in VE fluids, as opposed to HD fluids, facilitate the interaction between the widely placed droplets through the pairing of unlike sign inner vorticity lobes. In a mild-strong medium, the inner pair lobes remain in continuous vertical upward motion, while in a medium with strong coupling strength, the quicker-emerging waves are removing energy from the lobes at a faster rate. This reduces the rotating influence of the lobes more quickly; consequently, the upward-moving inner pair of density blobs begins to fill the gravitational supremacy and suddenly starts moving in a downward direction. In the case of medium-spaced: in VE fluids, similar to the HD scenario, the two new dipoles of unequal strength density blobs exhibit a circular motion and exchange the blobs. In mild strong fluid, the inner and outer density blobs move downward along the direction of gravity. However, in the medium of strong coupling strength, not like in the mild example, the inner blobs deform into filament shapes around the outer blobs. In the case of closely-spaced: the overlap of two opposite sign inner lobes reduces the impacts of each other. This results, the pair of two droplets shows the downward dynamics like a single pair under gravity. In mild, compared to the HD case, the wake-type structure is diminished, the lobes are better separated, and the vertical upward motion gets reduced. In strong, the faster shear waves result in slower lobe vertical propagation and greater horizontal spacing between the blobs.

This paper is organized in the following sections.  The next Section~\ref{model_meth} describes the basic model equations and the implementation of our numerical scheme. In Section \ref{simulation_results}, we report simulation results. The role of coupling strength on falling of droplet-droplet and rising of bubble-bubble density dynamics has been depicted in the form of TS waves through respective vorticity contour plots. To develop a better physical insight into the dynamics of each phenomenon, the inviscid limit of HD fluids is also simulated for each case. Final Section~\ref{conclusions}, conclude the paper with a summary.

\section{The numerical model and simulation methodology} 
\label{model_meth}
For a dusty plasma under gravity acceleration $\vec{g}$, the generalized hydrodynamic fluid model is given by a coupled set of continuity and momentum equations:
\begin{equation}\label{eq:continuity}
  \frac{\partial \rho_d}{\partial t} + \nabla \cdot
\left(\rho_d\vec{v}_d\right)=0{,}
  \end{equation}
 \begin{eqnarray}\label{eq:momentum0}
 &&\left[1+{\tau_m}\left(\frac{\partial}{\partial{t}}+{\vec{v}_d}\cdot \nabla\right)\right]\nonumber\\
 && \left[ {{\rho_d}\left(\frac{\partial{\vec{v}_d}}{\partial {t}}+{\vec{v}_d}{\cdot} \nabla{\vec{v}_d}\right)}+{\rho_d}\vec{g}+{\rho_c}\nabla \phi_{d} \right]\nonumber\\
 &&=\eta \nabla^2\vec{v}_d{,}
 \end{eqnarray}
respectively and the incompressible condition is given as
\begin{equation}\label{eq:incompressible}
 {\nabla}{\cdot}{\vec{v}_d}=0{.}
\end{equation}
This model describes a basic hydrodynamic fluid in the limit $\tau_m{=}0$ through the Navier-Stokes equation. We have thoroughly discussed the derivation of these normalized equations, as well as the procedure for their numerical implementation and validation, in our earlier works \citep{dharodi2014visco,dharodi2016sub}. Here, the number density of the dust fluid is $n_d$, the mass of the dust particle is $m_d$, and the mass density of the dust fluid is ${\rho_d}= {n_d}{m_d}$. The variables $\rho_c$, $\vec{v}_d$, and $\phi_d$, respectively, indicate the dust charge density, dust charge potential, and dust fluid velocity. The time, length, velocity, and potential are normalized using inverses of the dust plasma frequency $\omega^{-1}_{pd} = \left({4\pi (Z_d e)^{2}n_{d0}}/{m_{d0}}\right)^{-1/2}$, plasma Debye length $\lambda_{d} = \left({K_B T_i}/{4{\pi} {Z_d}{n_{d0}}{e^2}}\right)^{1/2}$, ${\lambda_d}{\omega_{pd}}$, and ${{Z_d}e}/{{K_B}{T_i}}$, respectively. The parameters $m_d$, $T_i$, and $K_B$ stand for the dust grain mass, ion temperature, and Boltzmann constant, respectively. The number density $n_d$ is normalized by the equilibrium value $n_{d0}$. We consider a constant charge on each dust grain, $Z_d$, which may be positive or negative. Dust particles are generally negatively charged, but in some circumstances—such as secondary electron emission \citep{shukla2015introduction} or afterglow plasma conditions—they can acquire a positive charge \citep{chaubey2021positive,chaubey2022coulomb,chaubey2022preservation}.

\subsection{Simulation methodology:} 
\label{num_methodology}
  For the numerical modeling the above generalized momentum equation (\ref{eq:momentum0}) is transformed into a set of two coupled equations,
\begin{eqnarray}\label{eq:vort_incomp1}
{{\rho_d}\left(\frac{\partial{\vec{v}_d}}{\partial {t}}+{\vec{v}_d}{\cdot} \nabla{\vec{v}_d}\right)}+{\rho_d}\vec{g}+{\rho_c}\nabla \phi_{d}={\vec \psi}
\end{eqnarray}
\begin{equation}\label{eq:psi_incomp1}
\frac{\partial {\vec \psi}} {\partial t}+\vec{v}_d \cdot \nabla{\vec \psi}=
{\frac{\eta}{\tau_m}}{\nabla^2}{\vec{v}_d }-{\frac{\vec \psi}{\tau_m}}{.}
\end{equation}
 The quantity ${\vec \psi}(x,y)$ represents the strain produced in the elastic medium by the time-varying velocity fields. We suppose that there is no beginning flow and the density gradient and potential gradient are taken along the y-axis. In opposition to the fluid density gradient, the acceleration $\vec g$ is applied. With small perturbations; density,  scalar potential, and dust velocity can be written as ${\rho_d}(x,y,t)={\rho_{d0}}(y,t=0)+{\rho_{d1}}(x,y,t)$, ${\phi_d}(x,y,t)={\phi_{d0}}(y,t=0)+{{\phi_{d1}}}(x,y,t)$ and $
	{\vec{v}_d}(x,y,t)=0+{\vec{v}_{d1}}(x,y,t)$, respectively. Under the equilibrium condition, $ {\rho_{d0}}{g} =-{\rho_c}{{\partial\phi_{d0}} /{\partial y}}$, rewrite Eq.~(\ref{eq:vort_incomp1}) using the perturbations. 
 \begin{eqnarray}\label{eq:vort_incomp2}
\frac{\partial{\vec{v}_d}}{\partial {t}}+{\vec{v}_d}{\cdot} \nabla{\vec{v}_d}+{\frac{\rho_{d1}}{\rho_d}}\vec{g}+{\frac{\rho_{c}}{\rho_d}}{\nabla{\phi_{d1}}}=\frac{\vec \psi}{\rho_d}{.}
 \end{eqnarray}
Using the Boussinesq approximation ($\rho_{d0}{\gg}\rho_{d1}$) and the curl of Eq.~(\ref{eq:vort_incomp2}), we obtain 
\begin{equation}\label{eq:vort_incomp3} 
\frac{\partial{\xi_{z}}} {\partial t}+\left(\vec{v}_d \cdot \vec \nabla\right)
{\xi_{z}}={\frac{1}{\rho_{d0}}}{{\nabla}{\times}{\rho_{d1}}{\vec{g}}}+{\nabla}{\times}{\frac{\vec \psi}{\rho_d}}{.} 
\end{equation}
Since we have considered  a constant charge medium so ${\nabla}{\times}{\nabla}{\phi_{d1}}=0$. Here, the vorticity $ {\xi_{z}}(x,y)={\vec \nabla}{\times}{\vec v_d}(x,y) $ is normalised with dust plasma frequency. The final numerical model equations in term of variables x and y become 
\begin{equation}\label{eq:cont_incomp4}
 \frac{\partial \rho_d }{\partial t} +  \left(\vec{v}_d\cdot
\nabla\right)\rho_d= 0{,}
    \end{equation}
\begin{equation}\label{eq:psi_incomp4}
\frac{\partial {\vec \psi}} {\partial t}+\left(\vec{v}_d \cdot \vec
\nabla\right)
{\vec \psi}={\frac{\eta}{\tau_m}}{\nabla^2}{\vec{v}_d }-{\frac{\vec
\psi}{\tau_m}}{,}  
\end{equation}
\begin{equation}\label{eq:vort_incomp4} 
\frac{\partial{\xi}_z} {\partial t}+\left(\vec{v}_d \cdot \vec \nabla\right)
{{\xi}_z}=-{\frac{g}{\rho_{d0}}}{\frac{\partial{\rho_{d1}}} {\partial x}}
+{\frac{\partial}{\partial x}}\left({\frac{\psi_{y}}{\rho_d}}\right)
-{\frac{\partial}{\partial y}}\left({\frac{\psi_{x}}{\rho_d}}\right){.}   
\end{equation}

We use the LCPFCT software \citep{boris_book} to solve the set of  above last three coupled equations numerically. The velocity is updated by using the velocity-vorticity relation ${\nabla^2}{\vec{v}_d}=-{\vec {\nabla}}{\times}{\vec \xi}$,  using the FISPACK \citep{swarztrauber1999fishpack}, at each time step. Boundary conditions in simulation studies are are non-periodic along the vertical (y-axis) direction, where the effects of perturbed values die out before hitting the simulation box boundary, but periodic in the horizontal (x-axis). To ensure grid independence of the numerical results in each example, a grid convergence analysis was conducted. In the HD limit $i.e.$~$\tau_m{=}0$,  the vorticity equation~(\ref{eq:psi_incomp4}) becomes
\begin{equation}\label{eq:vort_incomp_fluid} 
	\frac{\partial{\xi}_z} {\partial t}+\left(\vec{v}_d \cdot \vec \nabla\right)
	{{\xi}_z}=-{\frac{g}{\rho_{d0}}}{\frac{\partial{\rho_{d1}}} {\partial x}}+{\eta}{\nabla^2}{{\xi}_z}{.}   
\end{equation}
Therefore, we solve the set of equations~(\ref{eq:cont_incomp4}) and ~(\ref{eq:vort_incomp_fluid}) numerically for pure HD situations.  Using the previously mentioned velocity-vorticity relationship, the fluid velocity at each time step is updated. The same numerical model and simulation methodology, with some more details, were discussed in parts I and II. Note that whereas vorticity evolution has been described in terms of lobe/lobes, density evolution has been described in terms of blob/blobs. 
\section{Simulation results}\label{simulation_results}
The total density is $\rho_d=\rho_{d0}+\rho_{d1}$. ${\rho_{d0}}$ is the background density. The net Gaussian density inhomogeneity is given by 
 \begin{equation}
 \label{eq:dd0_profile} 
\rho_{d1}={\rho^{\prime}_{d1}+\rho^{\prime}_{d2}}={\rho^{\prime}_{01}}{{e^{-{r_1^2}/{a^2_{c1}}}}}+
  {\rho^{\prime}_{02}}{{e^{-{r_2^2}/{a^2_{c2}}}}}.
   \end{equation}
  Here $r_1^2={\left(x-x_{01}\right)^2+\left(y-y_{01}\right)^2}$ and $r_2^2={\left(x-x_{02}\right)^2+\left(y-y_{02}\right)^2}$. $a_{c1}$ and $a_{c2}$ are the initial radii of two droplets/bubbles. The seperation between two droplets/bubbles is given by a distance $d={x_{c2}}-{x_{c1}}$.  During the whole numerical simulations, ${\rho_{d0}}=5$, ${a_{c1}}={a_{c2}}={a_{c}}=2$ are kept constant. As a result, both the droplets/bubbles exhibit spatial symmetry. A system of length $lx=ly=48{\pi}$ units with $512{\times}512$ grid points in both the x and y axes has been taken into consideration. The system ranges from $-24{\pi}$ to $24{\pi}$ units along the x- and y-axes. 
  
  For droplets ${\rho^{\prime}_{01}}$=${\rho^{\prime}_{02}}$=0.5. Both droplets are placed side-by-side in a row at the same height, $({y_{01}}, {y_{02}})=(8{\pi},8{\pi})$. Three spacings between two droplets are simulated: widely spaced $(d=24{\gg}{2a_{c}})$, medium spaced $(d=6{>}{2a_{c}})$, and closely spaced $(d=4{=}{2a_{c}})$, as cases (i), (ii) and (iii), respectively. We have introduced the coupling strength as mild-strong ($\eta$=2.5, $\tau_m$=20), and strong ($\eta$=2.5, $\tau_m$=10) in each case.
\subsection{A pair of falling droplets}
\label{falling_droplets}
\paragraph{}
Prior to providing simulation findings, it would be beneficial to have some basic qualitative understanding of droplet-droplet dynamics. The vorticity equation~(\ref{eq:vort_incomp_fluid}) for an inviscid flow ($\eta$=$\tau_m$=0) using equation~(\ref{eq:dd0_profile}) becomes
\begin{equation}
\label{eq:fluid1} 
\frac{\partial{\xi_{z}}} {\partial t}+\left(\vec{v}_d \cdot \vec \nabla\right)
{\xi_{z}}={\frac{2g}{\rho_{d0}}}{\left(x-x_{c1}\right)}
{\rho^{\prime}_{d1}}+{\frac{2g}{\rho_{d0}}}{\left(x-x_{c2}\right)}{\rho^{\prime}_{d2}}{.}
\end{equation}
In this equation, the net vorticity of two droplets is represented by the RHS. Both terms describe falling dipolar vorticities under the action of gravity, each dipolar has two counter-rotating (or unlike-sign) lobes. These terms induce two dipoles (each have two lobes) to move downward by acting as a buoyant force. 

\begin{figure}
\centering              
\includegraphics[width=\textwidth]{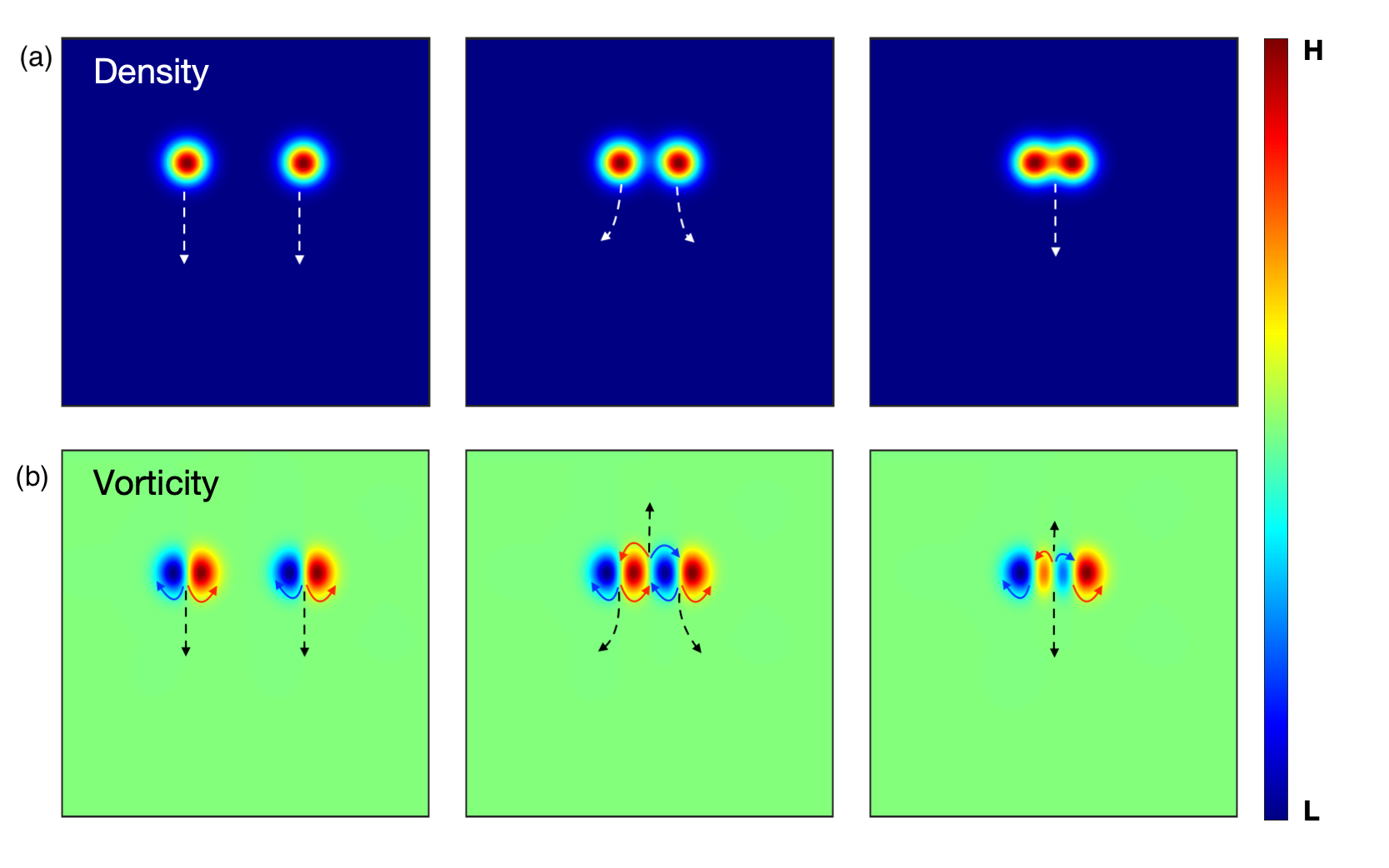}
\caption{A schematic diagram of a pair of droplets placed side by side in a row. The spacing between droplets decreases from left to right: (a) the density; and (b) the vorticity profile. The net propagation is indicated by the dotted arrows in (a,b) and the curved solid arrows represent the direction of rotation of the lobes in (b).  In the colourbar, The high-density region in the colorbar is represented by the abbreviation H, and the low-density/vorticicity region is represented by the letter L.}
\label{fig:figure1}
\end{figure}

Figure \ref{fig:figure1}(a), top row, shows the schematic profiles of two droplet densities positioned horizontally. For this configuration, bottom row (figure \ref{fig:figure1}(b)) displays the respective vorticities using equation~(\ref{eq:fluid1}). The curved solid arrows over the lobes represent their direction of rotation, while net propagation is indicated by the vertical and curved dotted arrows. In this schematic figure, the spacing between a two droplets decreases from left to right. Let's first discuss the case where the two droplets are widely spaced (left-top snapshot). The left-bottom snapshot depicts the corresponding counter-rotating vorticity lobes of each droplet. Here, the left lobe rotates clockwise while the right one is anti-clockwise, resulting in a vertical downward motion (indicated by the vertical downward dotted arrow). Due to the large spacing between them, there is essentially no contact between these two falling droplets; hence, gravity would mostly control their dynamics, just like in the individual examples. 

On the other hand, reducing the separation distance (refer to the remaining subplots from left to right) enhances the possibility of vorticity pairing through the counter-rotating inner lobes. This decreasing distance between two inner lobes, where the left lobe rotates anti-clockwise while the right one is clockwise, induces a new dipolar structure with vertical upward motion, as seen by the vertical dotted arrow in the middle snapshot of the bottom row. At the same time, these inner lobes pair with the outer lobes, which results in downward motion. The vorticity contour plot makes evident that along the gravity, the inner lobes compress and elongate in shape in comparison to the outer ones. As a result, both droplets begin to move apart by following a curved trajectory with an outward curvature instead of vertical fall. The curvature of this trajectory is determined to be proportional to the pairing of the vorticity lobes until they overlap. In the right snapshots from top and bottom rows, since the two droplets are nearly in contact with one another, the overlap of two opposite sign inner lobes eliminates each other's effects. As a result, the dipolar structure falls as one unit. The explanation makes clear that the competition between the vertical fall of two dipolar vorticities due to gravity and the interaction created by the pairing of inner vorticity lobes governs the overall dynamics. Under the influence of gravity, the dynamics described above are especially valid for an inviscid fluid.

In VE fluids, the vorticity equation (\ref{eq:vort_incomp3}) include an extra term ${\nabla}{\times}{{\vec \psi}/{\rho_d}}$ on the right-hand side of equation~(\ref{eq:fluid1}) in addition to the gravity term. This term refers to the TS waves that excite in the medium from the rotating lobes. The speed of these waves $(\sqrt{\eta/\tau_m})$ is proportional to the coupling strength $(\eta/\tau_m)$ of the medium. The ensuing subsections will provide a detailed visualization of the droplet-droplet dynamics using the numerical simulations. In simulations, for a fixed d, the effect of the VE nature on  the droplet–droplet interactions has been introduced through varying the coupling strength of the medium. 
\subsubsection{Widely spaced $(d=24>>2{a_{c}};{a_{c}}=2)$}
\label{widely_spaced_droplet}
\paragraph{}
Here, both the droplets $({x_{c1}}, {x_{c2}})=(-12, 12)$ are initially separated from one another by a considerable distance $(d=24>>2{a_{c}};{a_{c1}}={a_{c2}}={a_{c}}=2)$, with no overlap between the inner lobes of vorticities. 

\begin{figure}[h]
\centering              
\includegraphics[width=\textwidth]{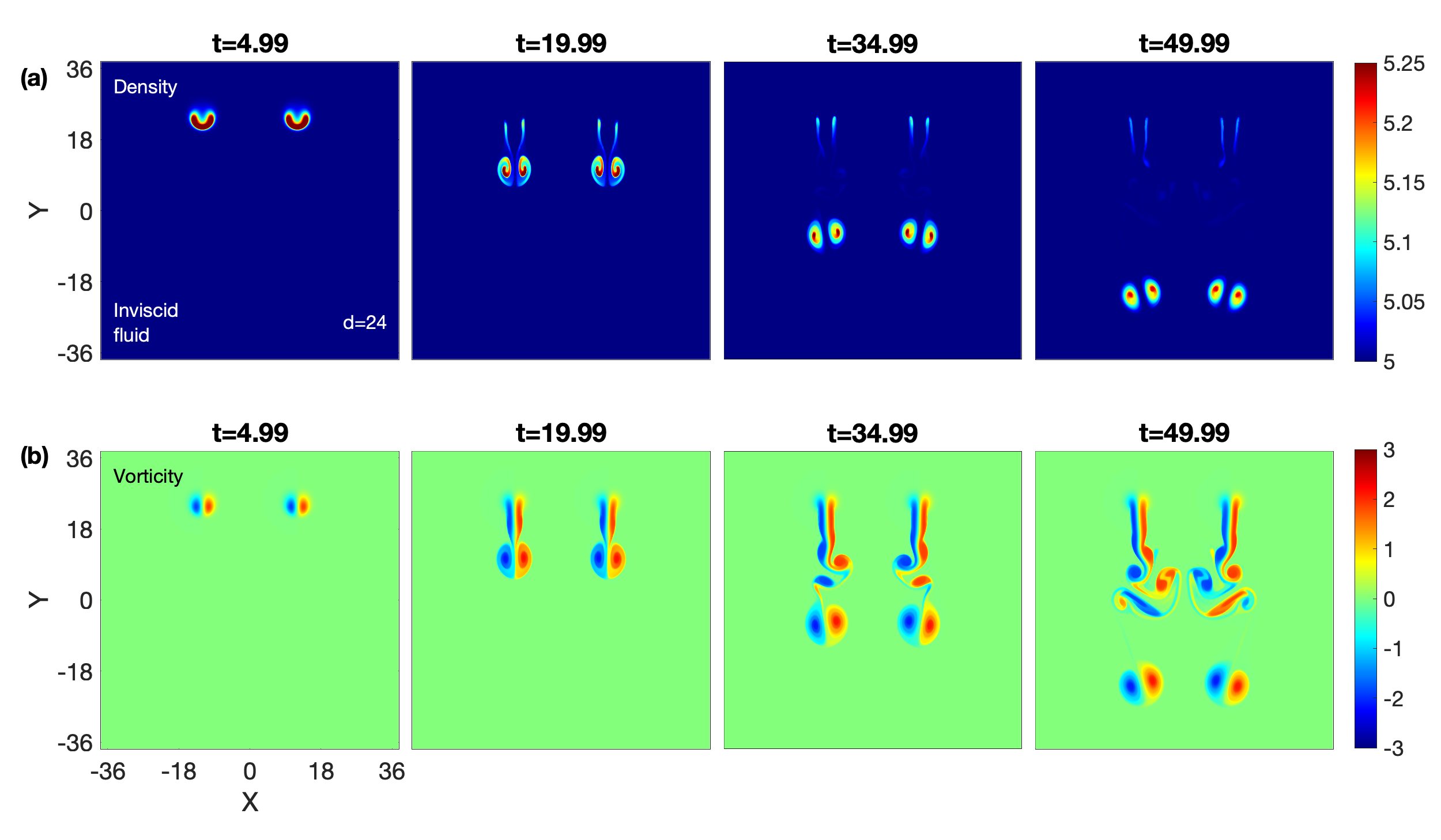}
\caption{Hydrodynamic inviscid fluid. Time evolution of two widely separated droplets density (a) (the colorbar designates the density) and vorticity (b) (the colorbar shows the vorticity)  in an inviscid fluid ($d{\gg}2{a_{c}}$), separated by distance $d = 24$ units. Both droplets have nearly independent dynamics and fall the same as the individual ones.}
\label{fig:figure2}
\end{figure}

Let's first examine the dynamics of this pair of droplets falling side by side in an inviscid HD fluid. Since there is no dissipation or source term except gravity, this combined evolution of each droplet is supposed to be similar as an individual one. The dynamics of an individual falling droplet were covered in Part I (ref. figure 14). In an HD system, the stages that occur when an individual droplet falls are as follows: the density blob, which were initially circular, take on crescent shapes due to gravity. Further, as time goes on, each droplet blob splits into two separate density blobs, or each droplet takes the dipolor form. These dipolor structures fall downward as a single entity and leave behind a wake-like structure in the background fluid. The time evolution of the combined density profile for both in figure \ref{fig:figure2}(a) makes these facts quite visible. The vorticity progression in figure~\ref{fig:figure2}(b) illustrates the cause of the falling droplet. At the onset of the simulation, the buoyant forces are induced in the form of dipolar vorticities for each droplet, each with two counter-rotating lobes. As a consequence, as was covered in more detail above, the droplet goes vertically downward.

\paragraph{}
 The droplet-droplet dynamics in SCDPs, which are modeled as VE fluids, is what we would like to see next. In SCDPs, in addition to the falling trajectory, each rotating vorticity lobe emits the shear waves into the ambient fluid at a speed that is proportional to the medium's coupling strength (${{\eta}/{\tau_m}}$). Stated differently, a medium possessing a higher coupling strength would be able to sustain the quicker TS waves. Lobes spread more quickly as a consequence of the quicker wave's ability to cover more distance in the same amount of time. In the following section, assuming the fixed viscosity $\eta=2.5$, the elastic factor $\tau_m$ solely modifies the coupling strength.

\begin{figure}
\centering              
\includegraphics[width=\textwidth]{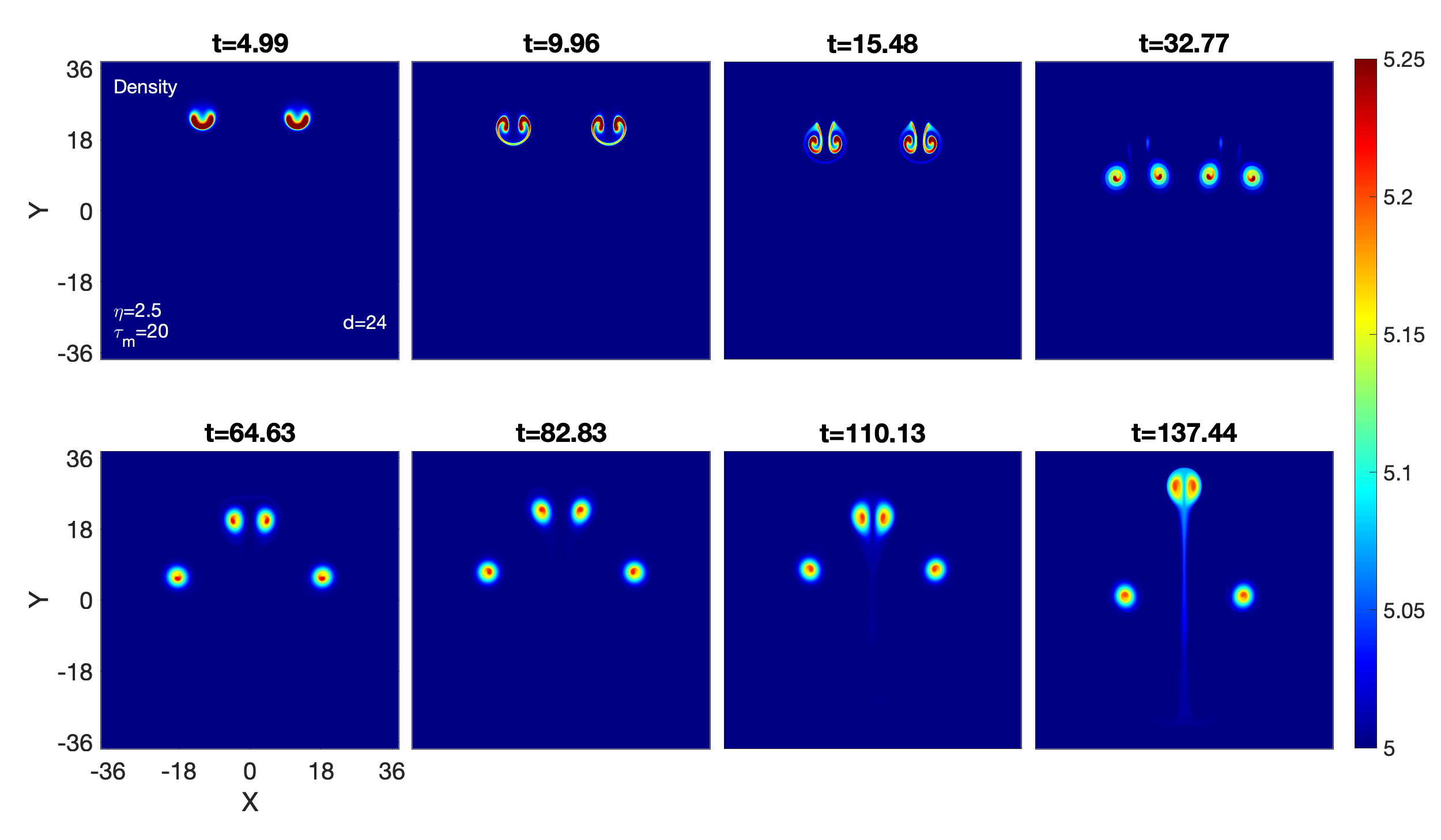}
\caption{Two widely separated droplets fall side by side over time in a SCDP/VE fluid with $\eta=2.5$ and $\tau_m=20$. The color bar indicates the density, which is common for all the subplots. Unlike HD fluid (see figure~\ref{fig:figure2}), we find that shear waves facilitate the pairing between two droplets.}
\label{fig:figure3}
\end{figure}

\begin{figure}
\centering               
\includegraphics[width=\textwidth]{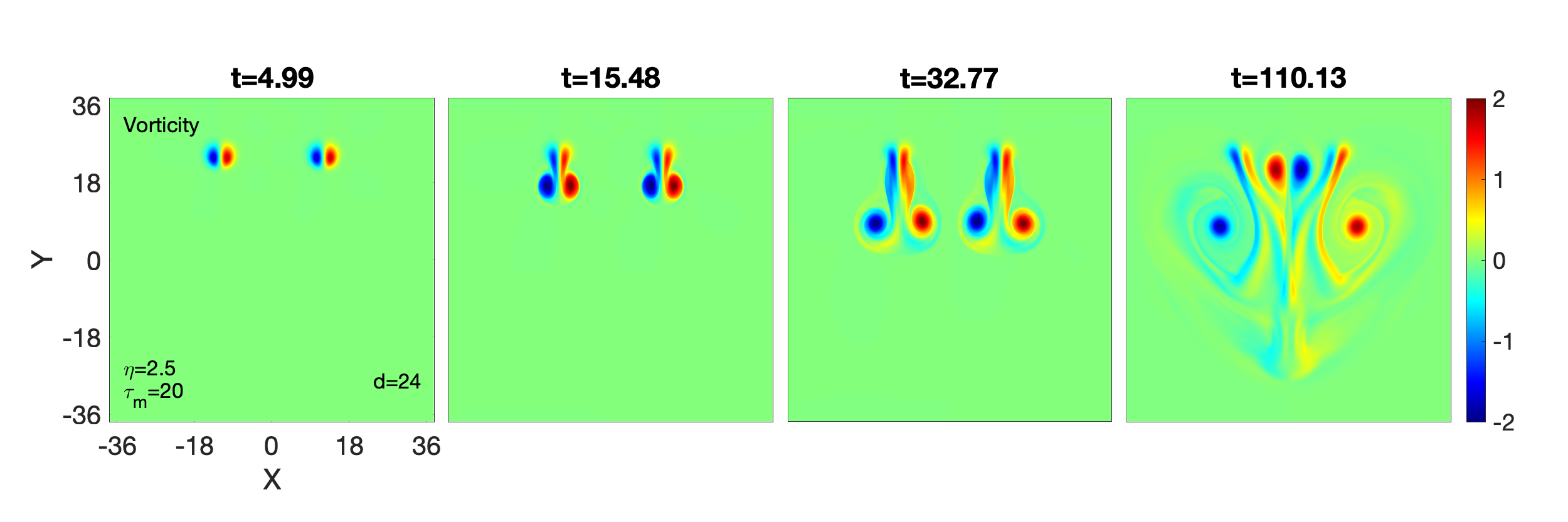}
\caption{The time evolution vorticity of two widely separated droplets falls side by side in a SCDP/VE fluid with $\eta=2.5$ and $\tau_m=20$ (see figure~\ref{fig:figure3} for the respective density evolution).}
\label{fig:figure4}
\end{figure}

Let's start with mild-strong VE fluid, with coupling parameter values of $\eta=2.5$ and $\tau_m$=20. This medium supports the emission of shear waves with the phase velocity $v_p=\sqrt{\eta/\tau_m}$ = 0.35 into the surrounding fluid from both lobes. For this medium, the density evolution is depicted in figure~\ref{fig:figure3}. Similar to the HD scenario (figure~\ref{fig:figure2}(a)), both of the circular density blobs in this instance first take on crescent shapes before splitting into two distinct blobs of density. Subplots, up to t=15.48, of the figure~\ref{fig:figure3} clearly display it. Further, as time goes on  (from t=32.77 to t=137.44), the inner blobs come close and start moving in a vertical upward direction as a single entity, while the outer start fall at a slower rate. The reason for this different dynamics can be understood from the vorticity progression in figure~\ref{fig:figure4}.
 
 In figure~\ref{fig:figure4}, the subplots at $t=32.77$ and $t=110.13$ clearly depict the shear waves originating from each vorticity lobe. Shear waves cause the counter-rotating lobes to begin pressing against one another perpendicular to the fall of the overall structure. Consequently, the lobes gradually separate and the inner lobes get pinched in between the outer lobes. The outcome is closer proximity between the inner counter-rotating lobes and greater separation between the outer lobes since these are free to move away. The closeness of the inner lobes, the left lobe rotates anti-clockwise while the right one clockwise, produces a new dipolar structure that is characterized by vertical upward motion. The leading (waves emitting from the outer lobes in the outer side) and pushing (waves emitting from the inner lobes) shear waves make the outer lobes propagate outward, which is perpendicular to the original propagation direction.  In addition, the strength of lobes/blobs is significantly diminished by the shear wave emission, which has been discussed by \cite{dharodi2016sub} and \cite{dharodi2024vortex}. These all contribute to slowing the fall of outer blobs. Here, it should also be highlighted that the shear waves have a greater effect on the inner blobs/lobes than the outside ones. As a result, the inner lobes/blobs have less strength and radial symmetry than the outside ones. 

\begin{figure}
\centering              
\includegraphics[width=\textwidth]{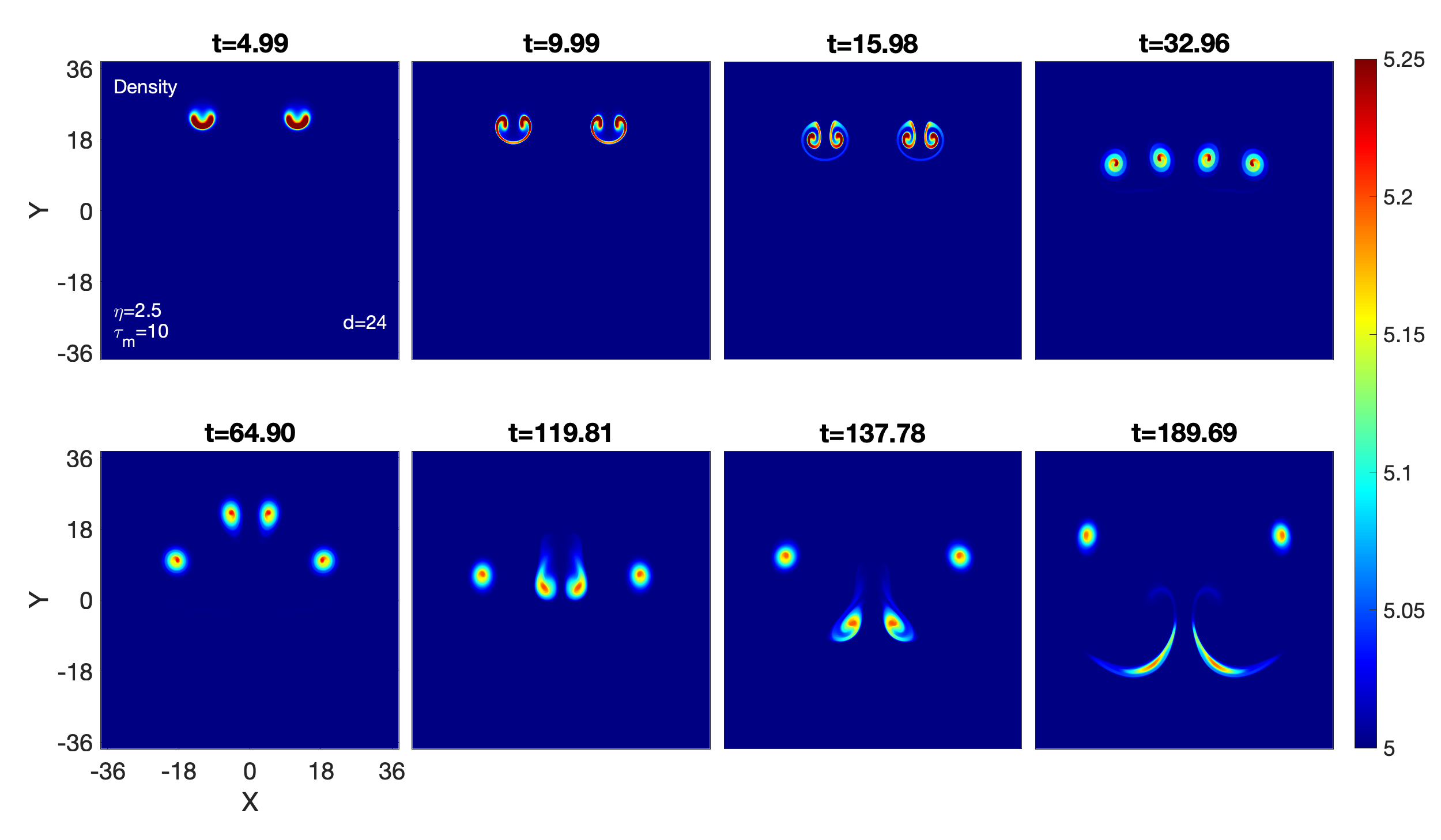}
\caption{In a SCDP/VE fluid with $\eta=2.5$; $\tau_m=10$, two widely separated droplets gradually fall side by side.}
\label{fig:figure5}
\end{figure}

\begin{figure}
\centering               
\includegraphics[width=\textwidth]{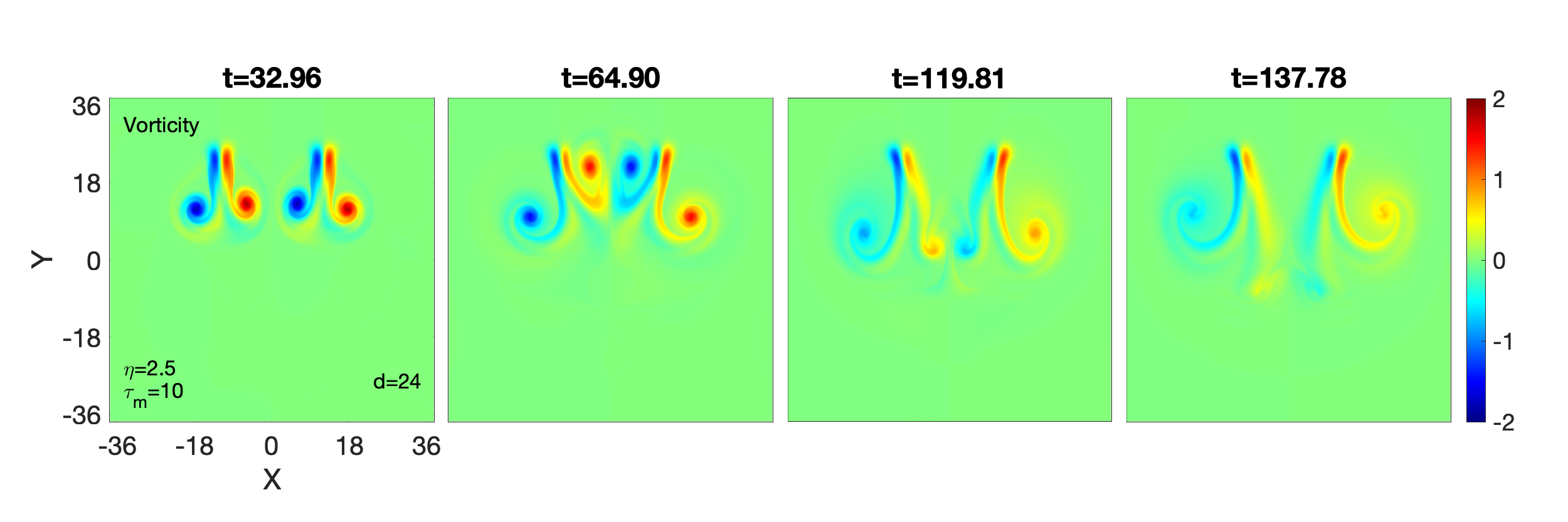}
\caption{The evolution vorticity of two widely separated droplets falls side by side  over time in a SCDP/VE fluid with $\eta=2.5$ and $\tau_m=10$ (see figure~\ref{fig:figure5} for the respective density evolution).}
\label{fig:figure6}
\end{figure}

Next, for the strong coupling strength $\eta=2.5$; $\tau_m=10$, figure \ref{fig:figure5} depicts the evolution of density, and figure \ref{fig:figure6} portrays the evolution of vorticity. Similar to the preceding example (figure \ref{fig:figure3}), the inner blobs in figure \ref{fig:figure5} come close to each other and begin to move in a vertical upward direction as a single entity. But, as time goes on, not like in the previous example, this inner pair suddenly starts moving in a downward direction. Eventually, the inner blobs deform into filament shapes around the upward moving outer blobs. It is distinctly shown in subplots 
from t=64.90 to t=189.69
of figure \ref{fig:figure5}.

 Figure~\ref{fig:figure6} shows the evolution of vorticity corresponding to the density evolution in  Fig.~\ref{fig:figure5}. Here, the shear waves moving at a phase velocity $v_p =\sqrt{\eta/\tau_m}$ = 0.5, which is greater than the above simulated scenario. The quicker-emerging waves removing energy from the lobes at a faster rate that results the lobes decay ultimately leads to its disappearance. This means, rotating influence of the lobes is reduced more quickly. Consequently, the upward-moving inner blob pair begins to fill the gravitational supremacy and suddenly starts moving in a downward direction.  Since the outer lobes are less affected by the emerging waves, they are supposed to remain symmetric and serve for a longer time. This results the inner lobes spiral around the rotating outer lobes.
\subsubsection{ Medium spaced $(d=6>2a_c, {a_{c}}=2)$}
\begin{figure}
\centering              
\includegraphics[width=\textwidth]{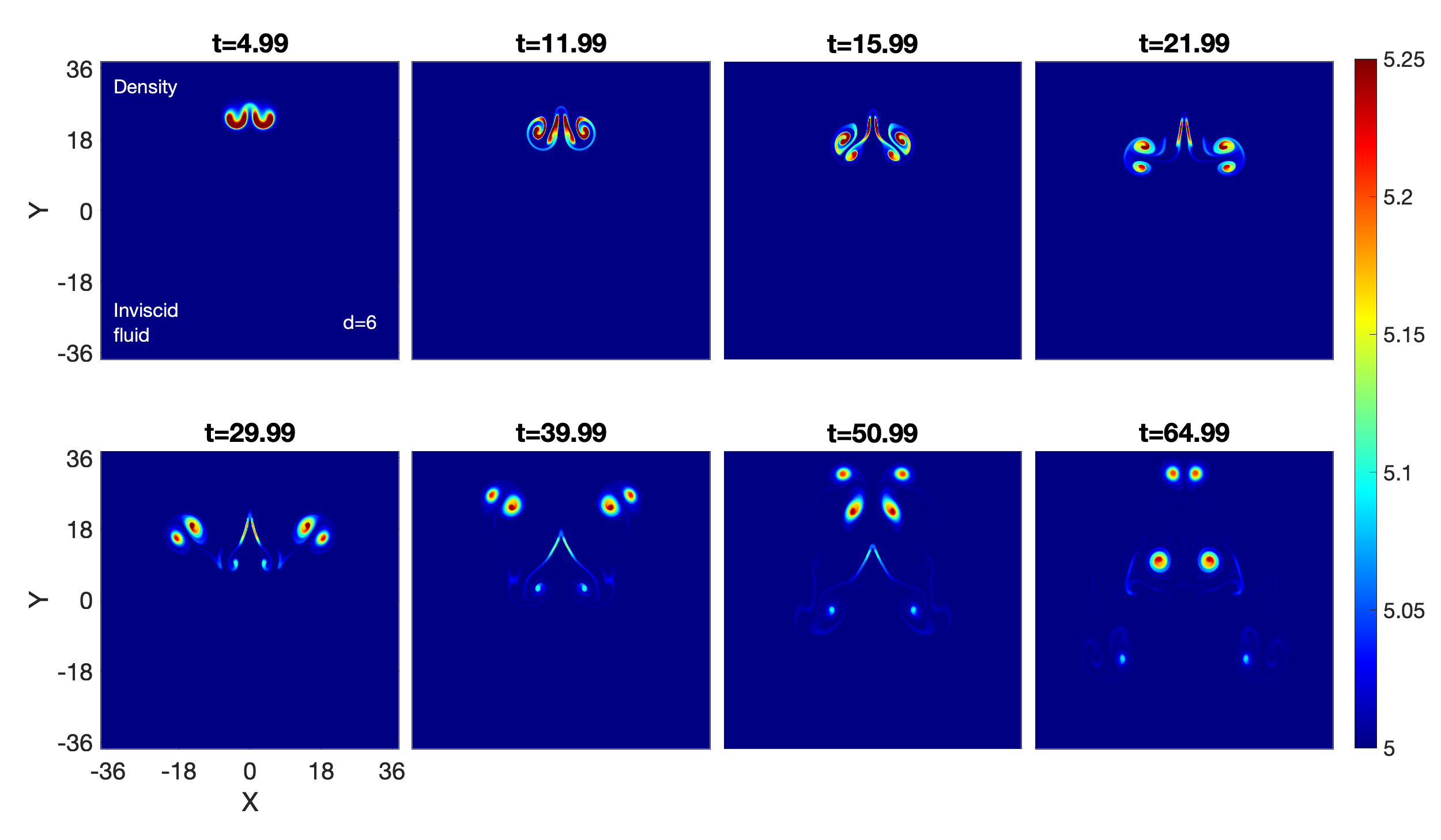}
\caption{Time evolution of the density of two medium-spaced droplets separated by $d = 6$ units in an inviscid fluid.}
\label{fig:figure7}
\end{figure}

\paragraph{}
Here, both the droplets $({x_{c1}}, {x_{c2}})=(-3, 3)$ are initially separated from one another by a distance $(d=6>2a_c, {a_{c}}=2)$, inner lobes of dipolar vorticities in close proximity to one another. Figure~\ref{fig:figure7} shows the evolution of density profile for this arrangement in inviscid HD fluid. Similar to the previous cases, after the crescent shapes, each density blobs split into two distinct blobs. 
However, because there is less space between blobs, the inner blobs become compressed and take on an elongated shape along the gravity. Under the influence of gravity, as inner blobs propagate downward, leaving behind a wake-like structure in the background fluid, their strength gets reduced more than the outer blobs. This results two new dipoles of unequal strength blobs. As a result, they propagate apart from one another and orthogonally to the original propagation direction. This makes the dipoles exhibit circular motion. The collision between the dipoles takes place which results the exchange of blobs. Now the outer blobs start propagating in the same direction of gravity while the inner in the opposite direction of gravity. During this period, the wake-like structure begins to travel downward under the force of gravity, further dividing into more dipolar blobs.

\begin{figure}
\centering               
\includegraphics[width=\textwidth]{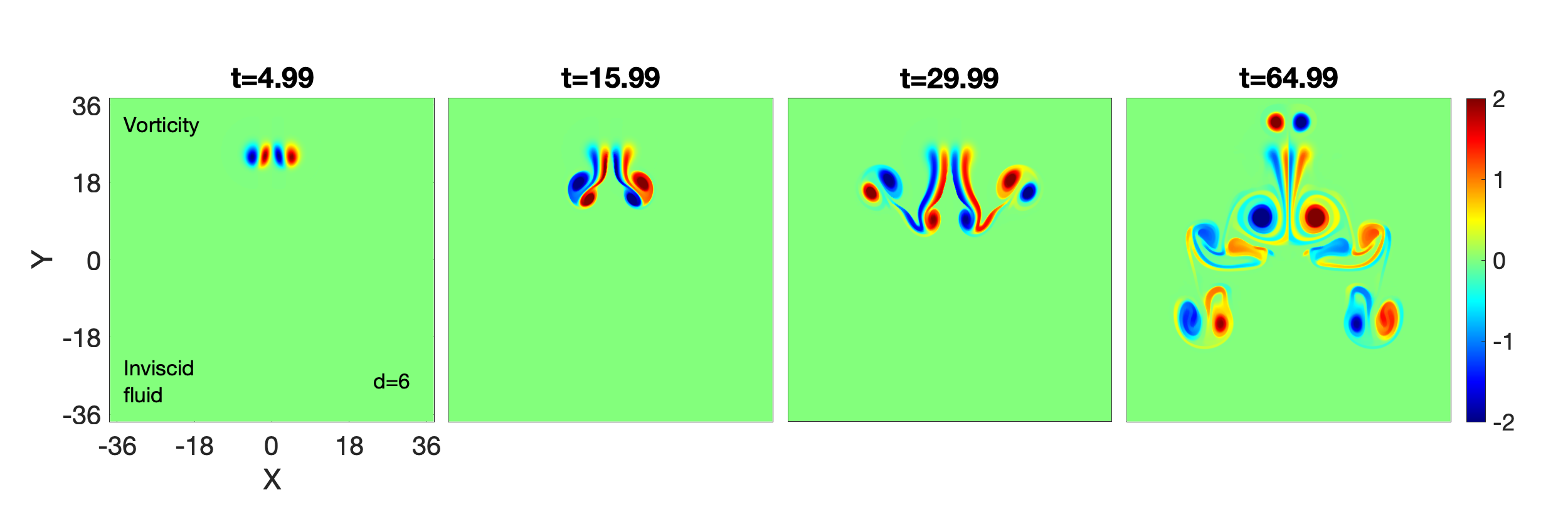}
\caption{Time evolution of the vorticity of two medium-spaced droplets in an inviscid fluid (see figure~\ref{fig:figure7} for the respective density evolution).}
\label{fig:figure8}
\end{figure}

Figure~\ref{fig:figure8} shows the evolution  vorticity  corresponding to the density evolution in  Fig.~\ref{fig:figure7}. 
For the main two dipoles, the inner weaker lobe try to rotate around the stronger outer lobe, that cause of the blobs/lobes moving away from one another in an orthogonal direction to the original propagation direction. Likewise, lobe formation occurs with wake-like structures.

\begin{figure}
\centering              
\includegraphics[width=\textwidth]{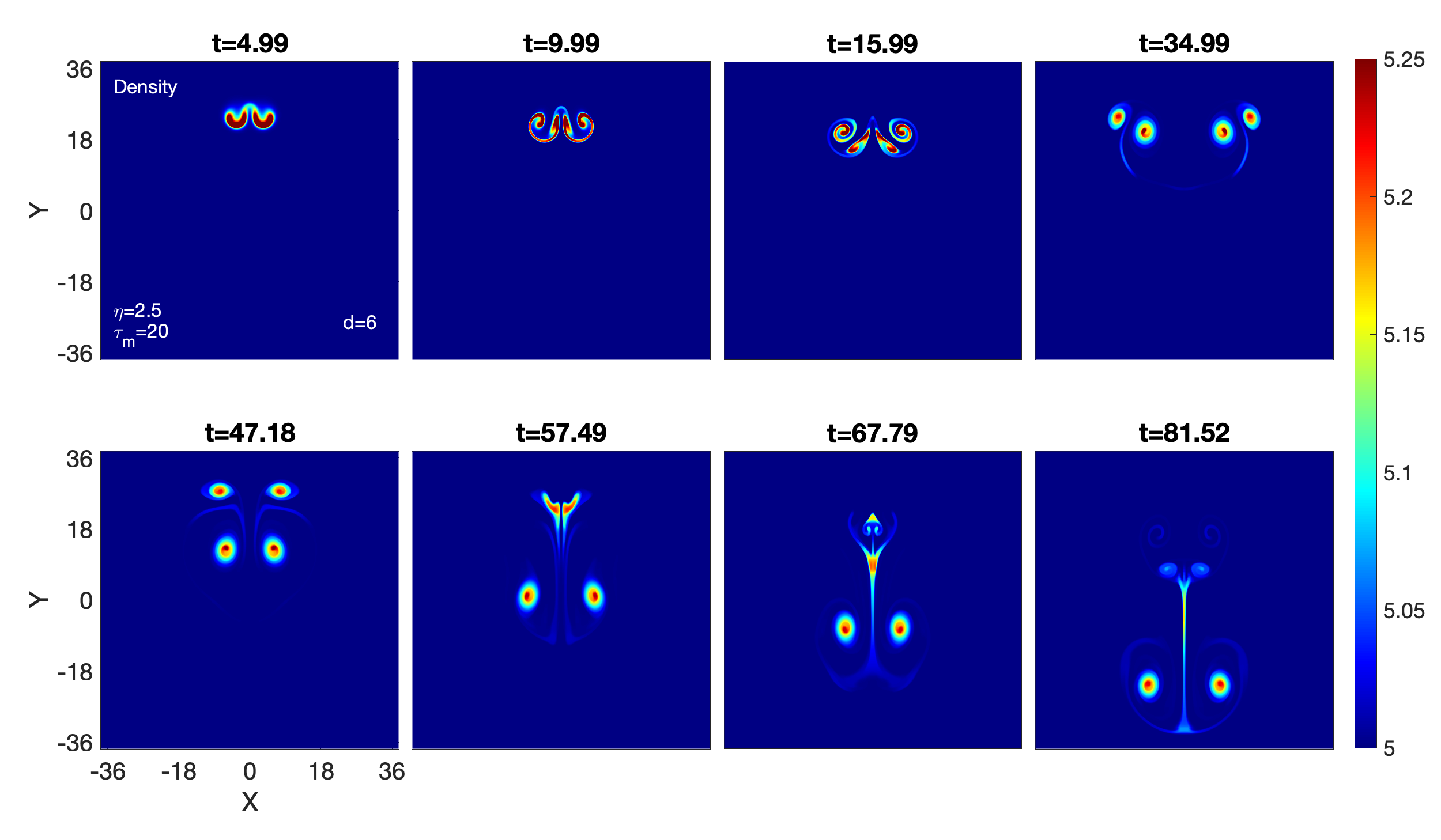}
\caption{Two widely separated droplets fall side by side over time in a SCDP/VE fluid with $\eta=2.5$ and $\tau_m=20$. The color bar indicates the density, which is common for all the subplots.}
\label{fig:figure9}
\end{figure}

Figure~\ref{fig:figure9}  shows the evolution of same density configuration for SCDP/VE case ($\eta$=2.5; $\tau_m$=20). Similar to the HD scenario (figure~\ref{fig:figure7}, both of the circular density blobs transform into two new dipoles of unequal strength blobs and exhibit a circular motion.  The collision between the dipoles results the exchange of blobs.  But here both the outer and inner blobs start move downward along the direction of gravity. Next to understand this dynamics let us discuss the vorticity progression. 

\begin{figure}
\centering               
\includegraphics[width=\textwidth]{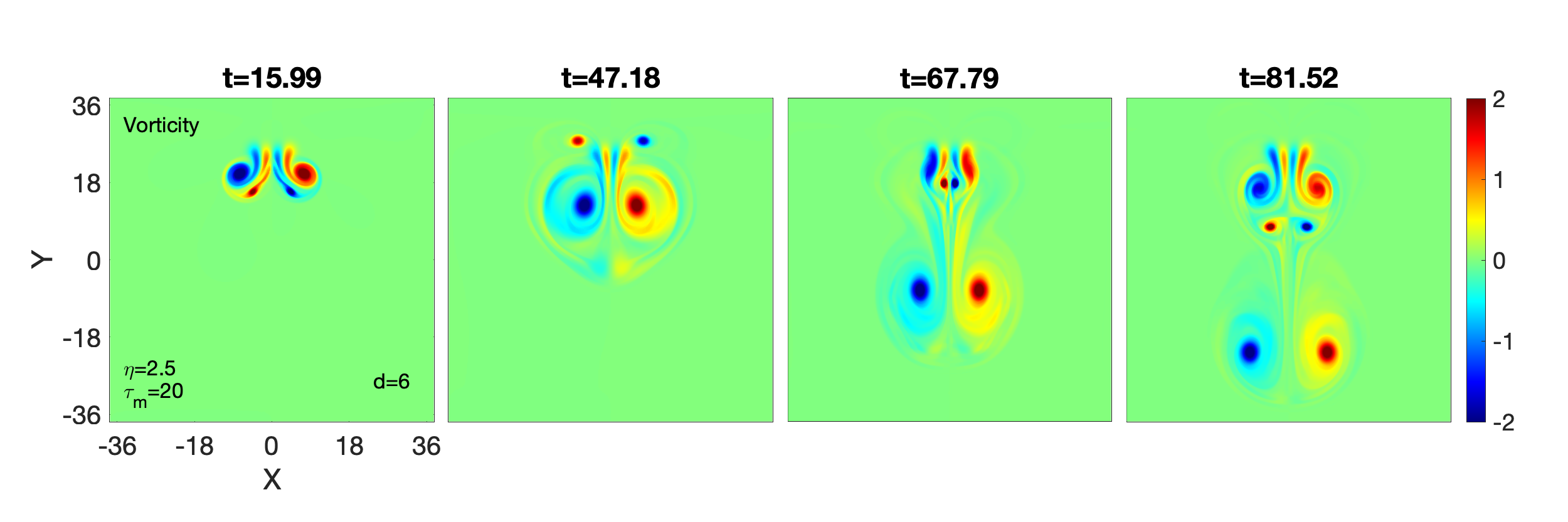}
\caption{The time evolution vorticity of two medium separated droplets falls side by side in a SCDP/VE fluid with $\eta=2.5$ and $\tau_m=20$ (see figure~\ref{fig:figure9} for the respective density evolution).}
\label{fig:figure10}
\end{figure}

Figure~\ref{fig:figure10} shows the evolution  vorticity  corresponding to the density evolution in  Fig.~\ref{fig:figure9}. This medium supports the shear waves moving with the phase velocity $v_p=\sqrt{\eta/\tau_m}$ = 0.35. The inner lobes are compressed in between the outer lobes by the shear waves that emanate from each rotating lobe. As a result, both dipoles split into two new dipoles, which move in a circle, with the weaker inner lobe trying to spin around the stronger outer lobe. The collision between the dipoles results the exchange of lobes. The dipolar structure formed due to the  outer (stronger) lobes continues to fall under gravity. At the time, the dipolar structure formed from the inner (weaker) lobes is supposed to travel in an upward direction, but this structure is engulfed by the shear waves, which behave like outer lobes for weaker lobes. Once again, this new arrangement produced two dipolar structures with unequal-strength lobes that moved in a circle and collided. As before, this configuration created two dipolar structures of unequal lobes which  performed a circular motion.

\begin{figure}
\centering              
\includegraphics[width=\textwidth]{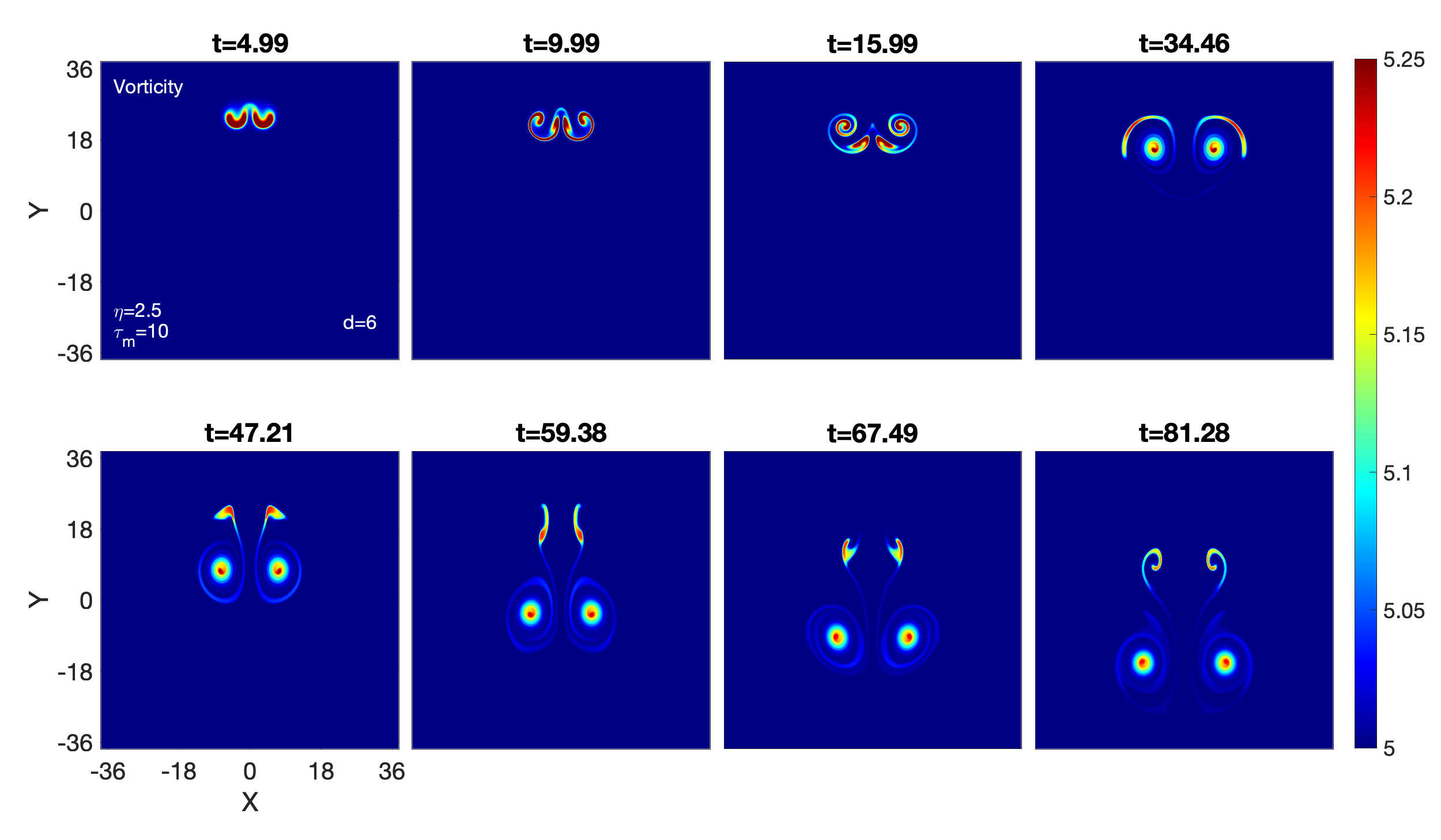}
\caption{Two widely separated droplets fall side by side over time in a SCDP/VE fluid with $\eta=2.5$ and $\tau_m=10$. The color bar indicates the density, which is common for all the subplots.}
\label{fig:figure11}
\end{figure}

\begin{figure}
\centering               
\includegraphics[width=\textwidth]{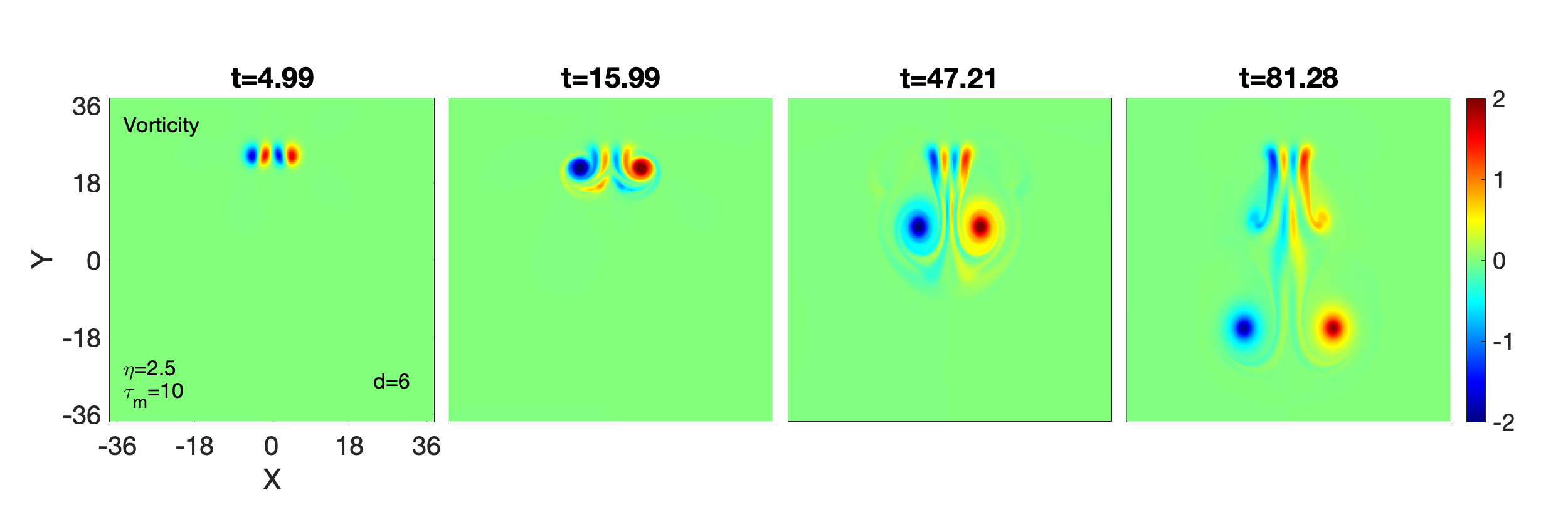}
\caption{The time evolution vorticity of two medium separated droplets falls side by side in a SCDP/VE fluid with $\eta=2.5$ and $\tau_m=10$ (see figure~\ref{fig:figure11} for the respective density evolution).}
\label{fig:figure12}
\end{figure}

Next, figure \ref{fig:figure11} shows the evolution of density for the strong coupling strength $\eta=2.5$; $\tau_m=10$, and figure \ref{fig:figure12} shows the evolution of vorticity. Similar to the preceding example (figure \ref{fig:figure9}), the two new dipoles of unequal strength blobs exhibit a circular motion in figure \ref{fig:figure11}. Not like in the previous example, the inner blobs deform into filament shapes around the outer blobs. It is distinctly shown in subplots from t=47.21 to t=81.28 of figure \ref{fig:figure11}.

 Figure~\ref{fig:figure12} shows the evolution of vorticity corresponding to the density evolution in  Fig.~\ref{fig:figure11}. Here, the shear waves moving at a phase velocity $v_p =\sqrt{\eta/\tau_m}$ = 0.5, which is greater than the above simulated scenario. Compared to the previous case, here, due to the quicker-emerging waves, the weaker inner blobs notice more compression and transform into spiral form around the outer (stronger) lobes at an earlier time. Thus, under the influence of gravity, only the outer lobe pair falls as a single entity.

\subsubsection{ Closely spaced $(d{\approx}{2a_{c}})$}
\begin{figure}
\centering              
\includegraphics[width=\textwidth]{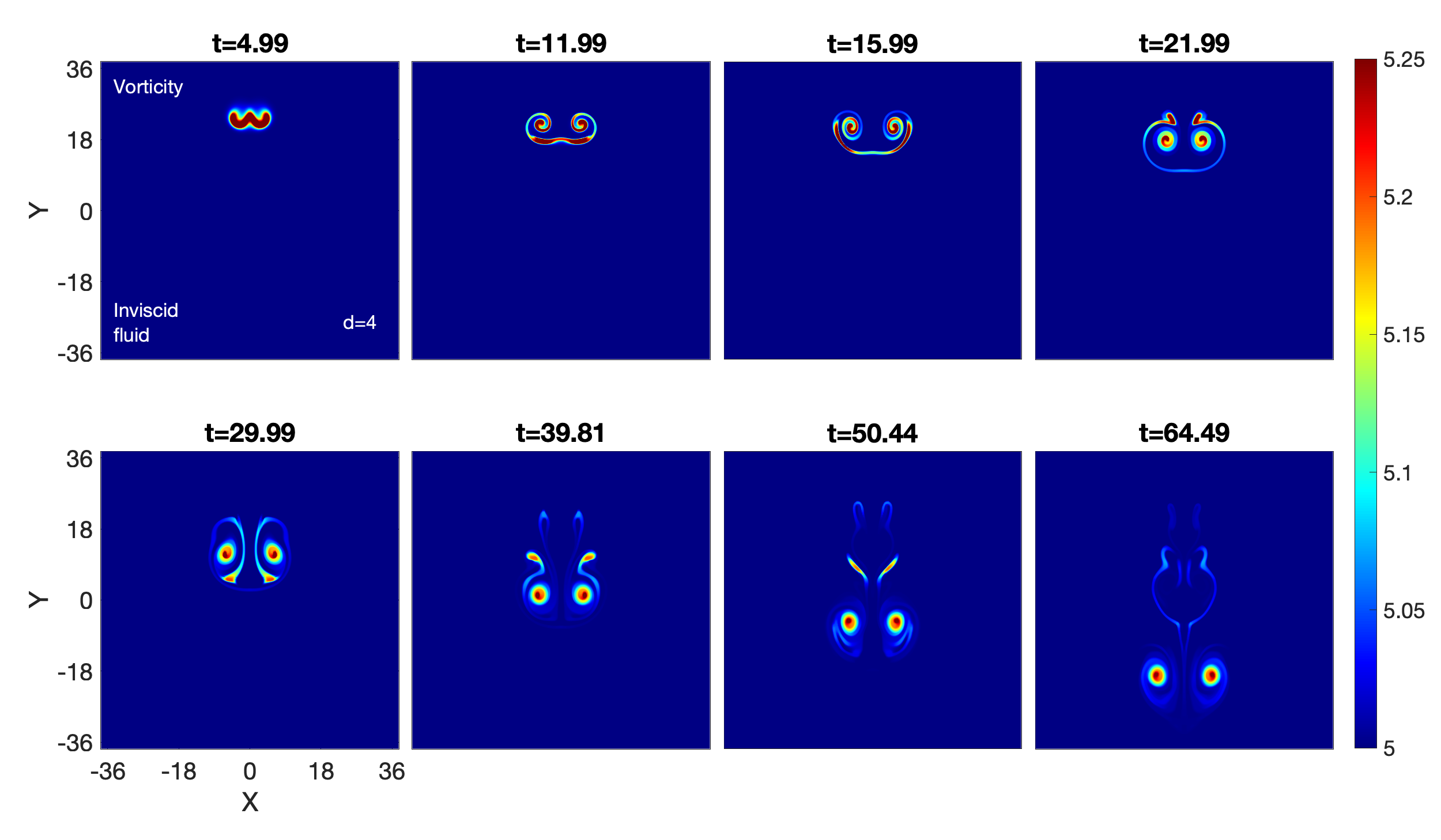}
\caption{Time evolution of the density of two closely-spaced droplets separated by $d = 4$ units in an inviscid fluid.}
\label{fig:figure13}
\end{figure}

\paragraph{}
To begin with, the droplet (${x_{c1}}=-2$) and bubble (${x_{c2}}=2$) are placed close enough $(d=4\approx2a_c)$ that both the like-sign inner lobes of vorticities almost overlap with each other. Figure~\ref{fig:figure13} shows the evolution of density profile for this arrangement in inviscid HD fluid. At the beginning of the simulation, the density produces a double-well-type structure, or "w". Over time, the central bump of this structure spirals around its rotating edges. This process keeps going while the dipole (outer lobes) falls because of gravity.

\begin{figure}
\centering               
\includegraphics[width=\textwidth]{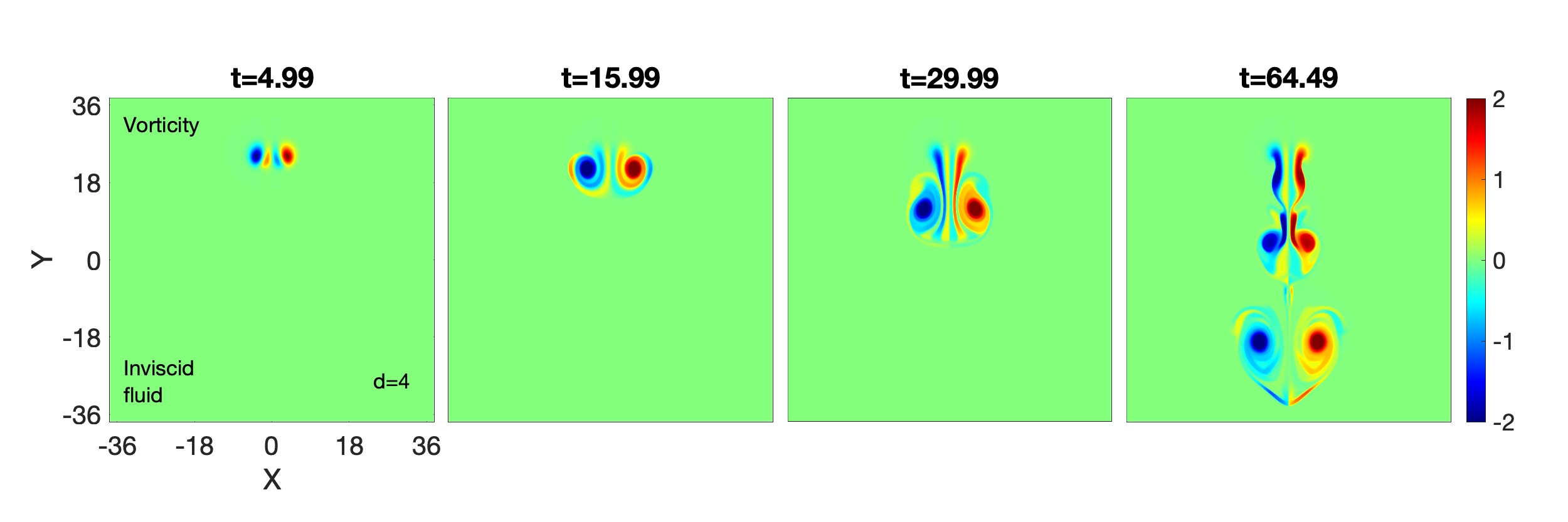}
\caption{Time evolution of the vorticity of two closely-spaced droplets in an inviscid fluid (see figure~\ref{fig:figure13} for the respective density evolution).}
\label{fig:figure14}
\end{figure}

Figure~\ref{fig:figure14} shows the evolution  vorticity  corresponding to the density evolution in  Fig.~\ref{fig:figure13}. The overlap of two opposite sign inner lobes reduces the impacts of each other. Consequently, the inner lobes merely wrap around the outer lobes, which play a major role in the creation of this dipolar structure. The inner lobes spiral around their revolving outer lobes throughout time. This process continues with the dipole falls because of gravity.

\begin{figure}
\centering              
\includegraphics[width=\textwidth]{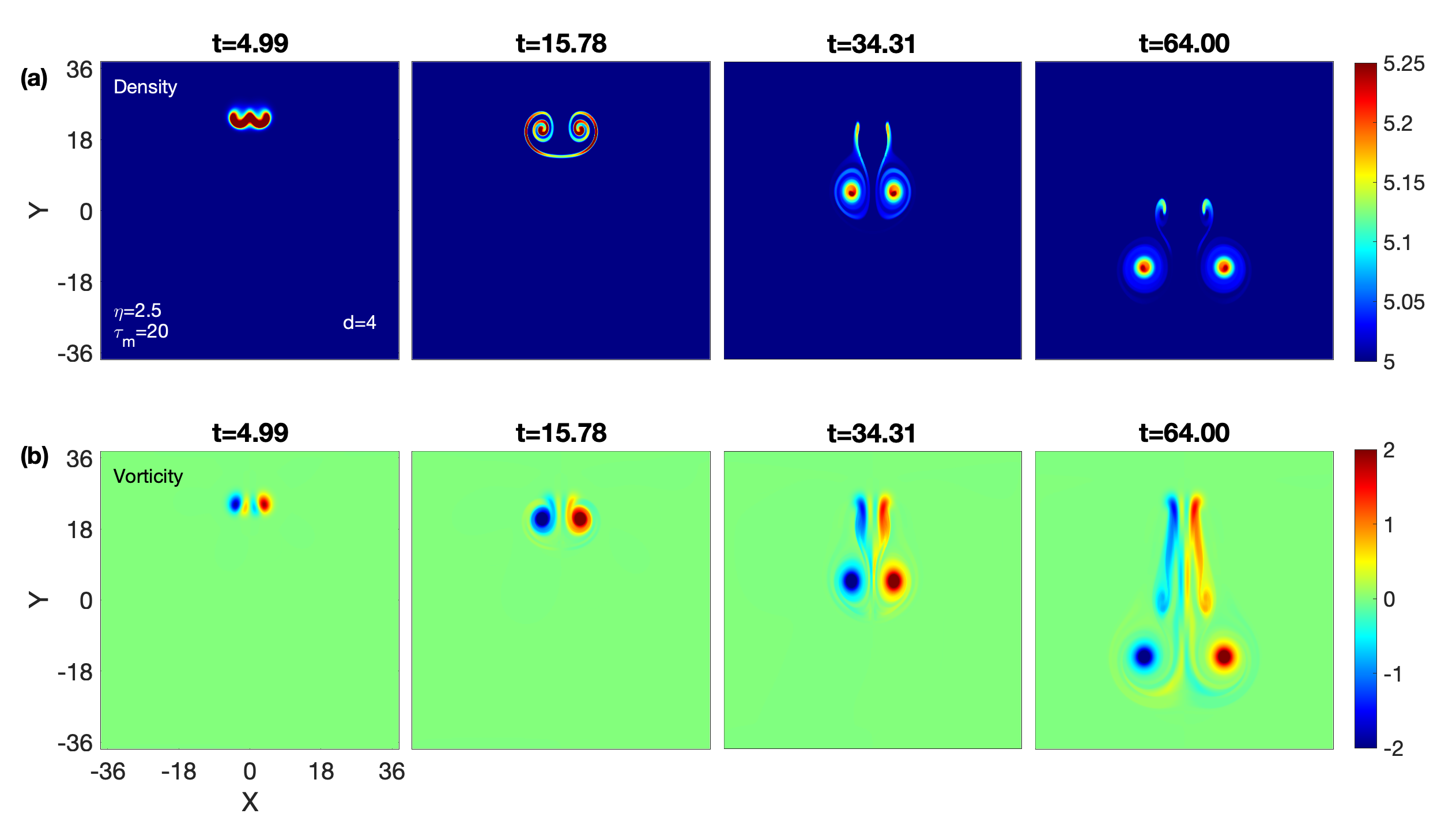}
\caption{
Time evolution of closely-spaced droplet–droplet density (a) and vorticity (b) in a SCDP/VE fluid with $\eta=2.5$ and $\tau_m=20$.}
\label{fig:figure15}
\end{figure}

Figure~\ref{fig:figure15}(a)  shows the evolution of same density configuration for SCDP/VE case ($\eta$=2.5; $\tau_m$=20). Similar to the HD scenario (figure~\ref{fig:figure13}), during the process of dipole falls, the central bump of the double-well or "w" type density profile spirals around its rotating edges. However, compared to the HD case, the wake-type structure is diminished, the lobes are better separated, and the vertical upward motion gets reduced.  These different observations in this VE fluid compared to basic HD fluid are because VE fluid favors the emission of waves from individual lobes.

\begin{figure}
\centering              
\includegraphics[width=\textwidth]{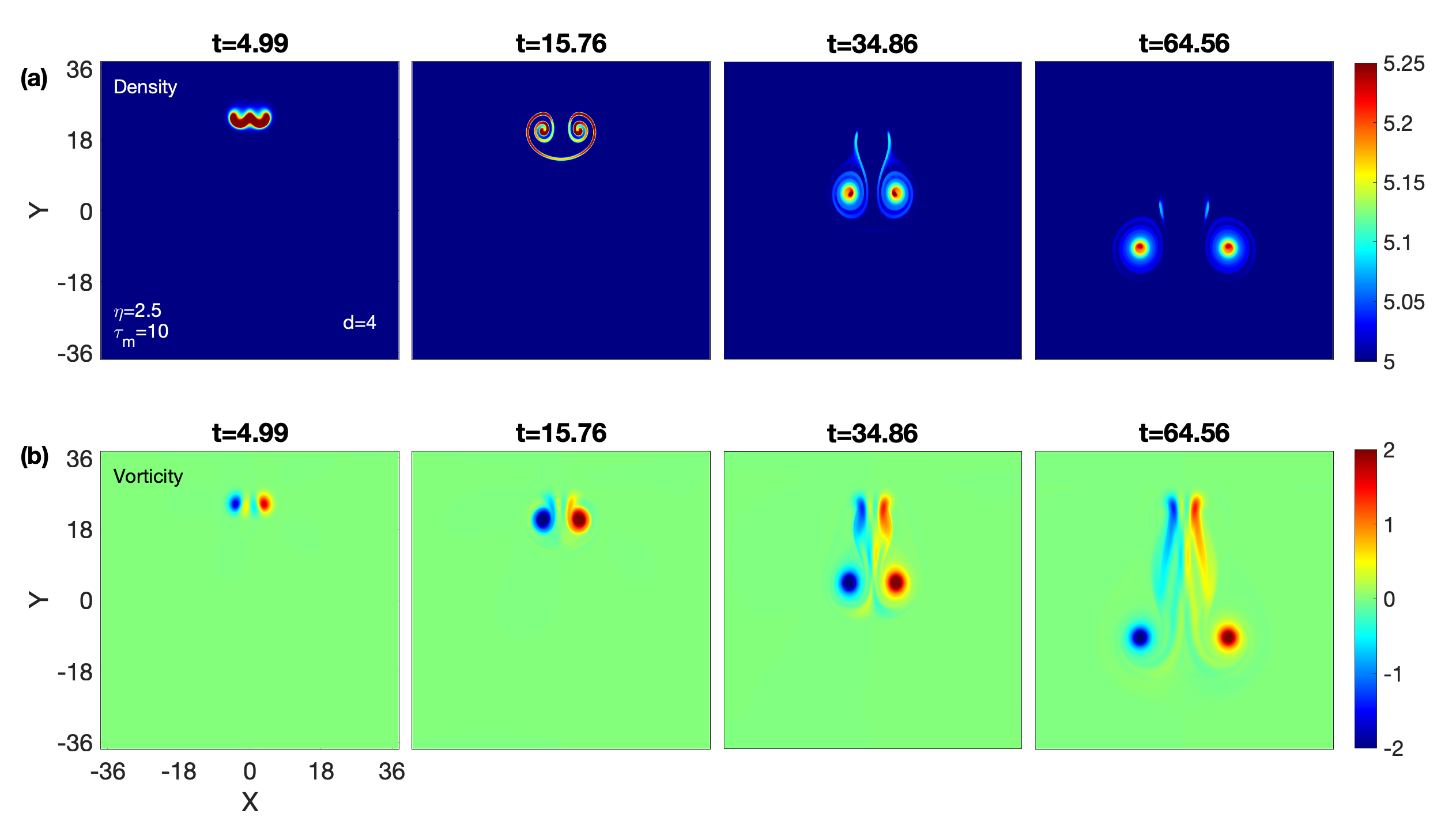}
\caption{
Time evolution of closely-spaced droplet–droplet density (a) and vorticity (b) in a SCDP/VE fluid with $\eta=2.5$ and $\tau_m=10$.}
\label{fig:figure16}
\end{figure}

A new VE fluid with $\eta = 2.5$ and $\tau_m = 10$ is simulated in figure \ref{fig:figure16}. This medium supports the shear waves moving at a phase velocity $v_p =\sqrt{\eta/\tau_m}$ = 0.5, which is greater than the above simulated scenario ($v_p =\sqrt{\eta/\tau_m}$ = 0.35; see figure \ref{fig:figure15}). The faster shear waves (or the higher coupling strength) result in slower lobe vertical propagation and greater horizontal spacing between the blobs. This can be observed by the comparison analysis of the two VE fluids.

\subsection{A pair of rising bubbles}
\label{rising_bubbles}
\paragraph{}

\begin{figure}
\centering              
\includegraphics[width=\textwidth]{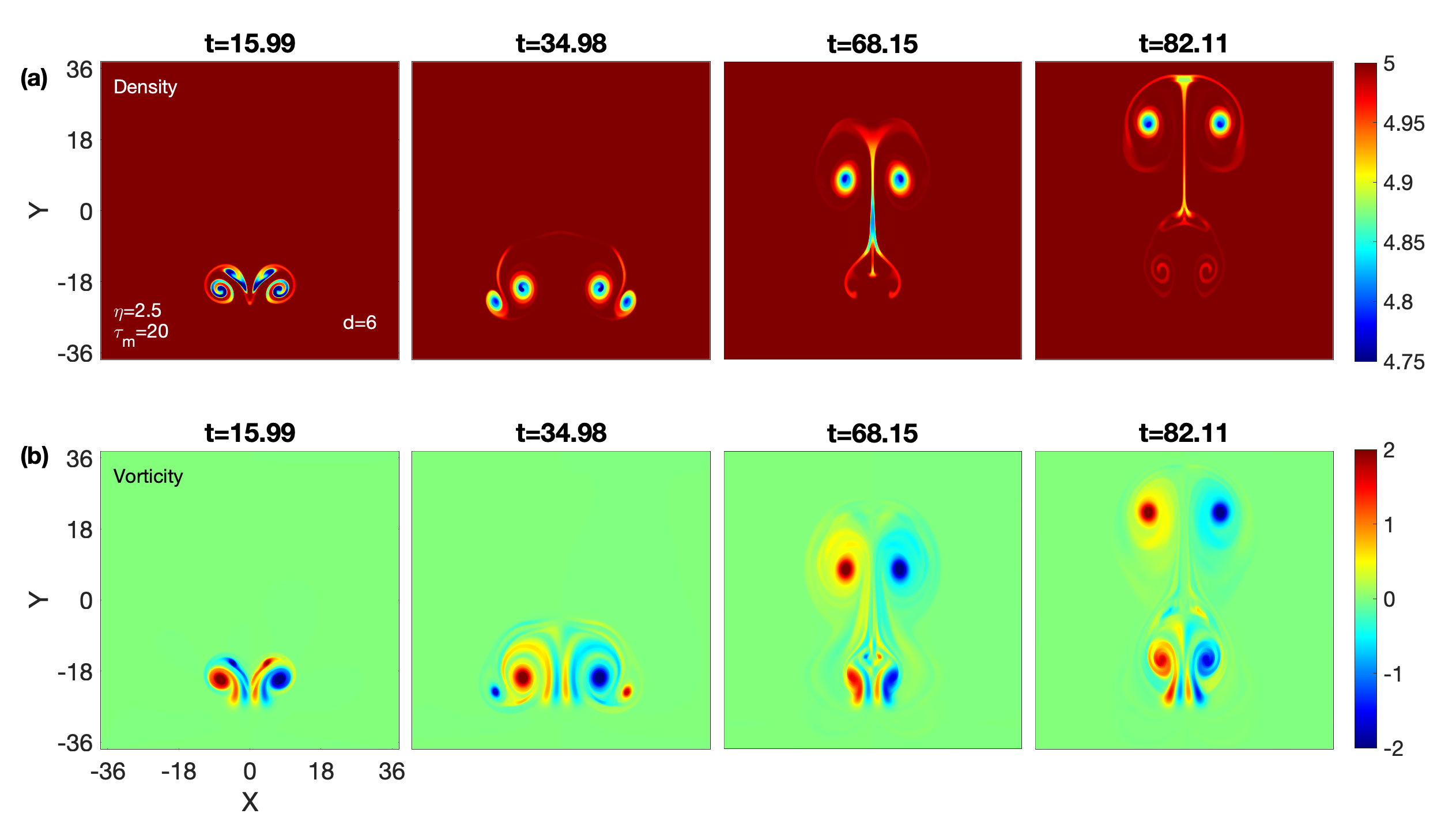}
\caption{
Time evolution of medium-spaced bubble–bubble density (a) and vorticity (b) in a SCDP/VE fluid with $\eta=2.5$ and $\tau_m=20$.}
\label{fig:figure17}
\end{figure}

 For the rising bubbles, we have taken into consideration that ${\rho^{\prime}_{01}}$=${\rho^{\prime}_{02}}$=-0.5. In Part I, we looked at a rising bubble and a falling droplet independently. We found that a rising bubble is analogous to the falling droplet process. Similarly, in the present study in each of the three cases, we have observed that the behavior of two falling droplets is the same as the behavior of two rising bubbles. Thus, the only case we cover is one in which two bubbles are medium-spaced $(d=6{>}{2a_{c}})$, where $({x_{c1}}, {x_{c2}})=(-3, 3)$ and $({y_{01}}, {y_{02}})=(-8{\pi},-8{\pi})$. Figure~\ref{fig:figure17} shows this density configuration for SCDP/VE case ($\eta$=2.5; $\tau_m$=20). Here, the two circular density blobs change into two new dipoles of unequal strength. These dipoles exhibit a circular motion and after collision results the exchange of blobs. Both the outer and inner blobs start moving upward in the opposite direction of gravity. A comparison of figures ~\ref{fig:figure17}(a) and ~\ref{fig:figure9} demonstrates that the process of two rising bubbles and two falling droplets are identical.

 Figure~\ref{fig:figure17}(b) shows the evolution  vorticity  corresponding to the density evolution in  Fig.~\ref{fig:figure17}(a). This medium favours the shear waves moving with the phase velocity $v_p=\sqrt{\eta/\tau_m}$ = 0.35. Due to the emerging shear waves from each lobe, two new dipoles formed, which moved in a circle. The weaker inner lobe attempt to rotate around the stronger outer lobe is what drives this circulation motion. The collision between the dipoles results the exchange of lobes. The outer (stronger) lobes continue to rise as a dipolar structure. The dipolar structure formed from the inner (weaker) lobes is swallowed by the shear waves, which behave like outer lobes for weaker lobes. Once again, this new configuration resulted in two dipolar structures with different lobe strength that moved in a circle and collided. The evolution of vorticity of two rising bubbles and two falling droplets are identical, as can be shown by comparing figures ~\ref{fig:figure17}(b) and ~\ref{fig:figure10}.

\begin{figure}
\centering              
\includegraphics[width=\textwidth]{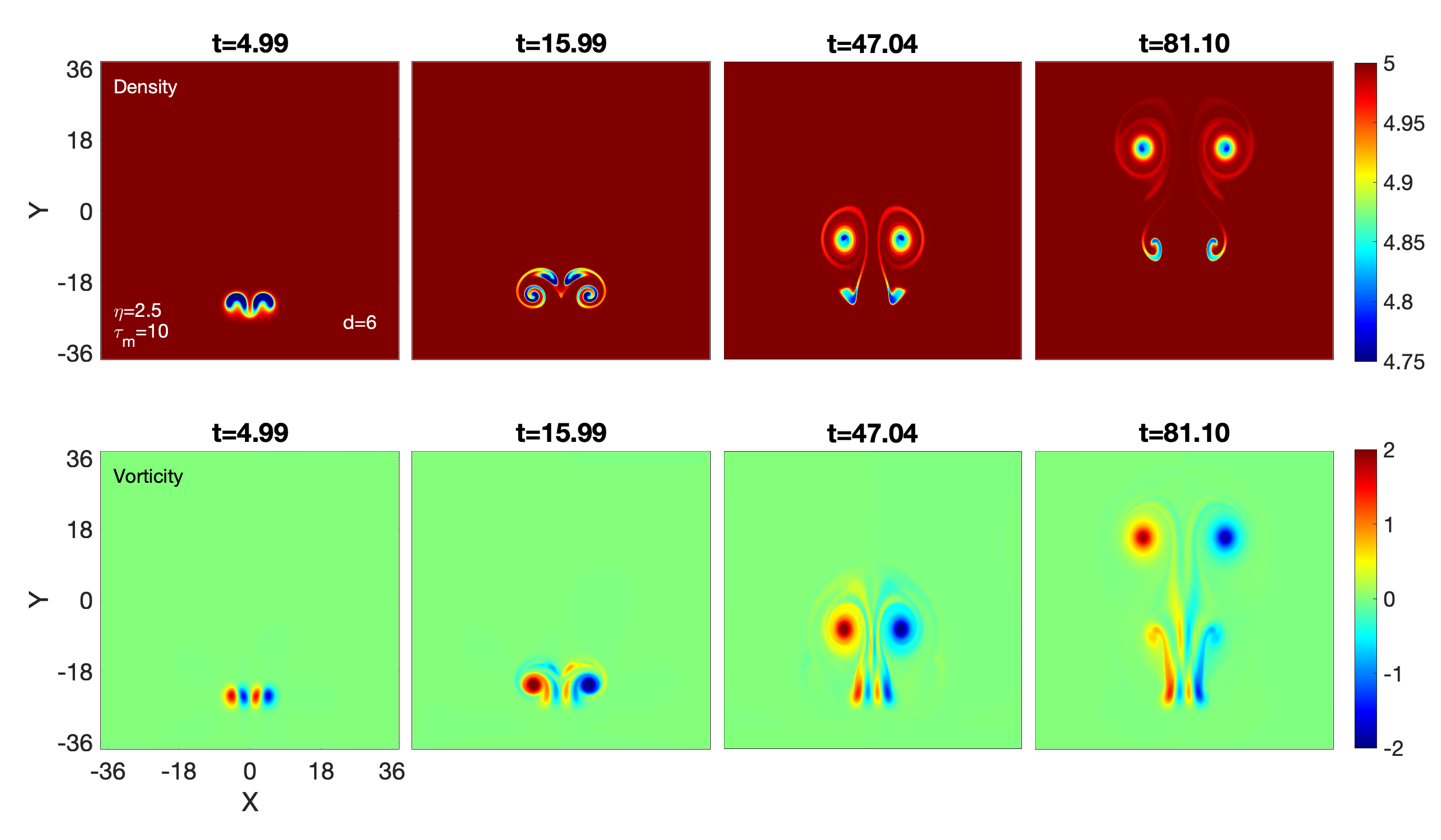}
\caption{
Time evolution of medium-spaced bubble–bubble density (a) and vorticity (b) in a SCDP/VE fluid with $\eta=2.5$ and $\tau_m=10$.}
\label{fig:figure18}
\end{figure}

Next, figure \ref{fig:figure18}(a) shows the evolution of density for the strong coupling strength $\eta=2.5$; $\tau_m=10$, and figure \ref{fig:figure18}(b) shows the evolution of vorticity. In this case, the evolution of vorticity/density profiles of two rising bubbles and two falling droplets are identical, as can be shown by comparing figures ~\ref{fig:figure18}(a) and ~\ref{fig:figure11}, and figures ~\ref{fig:figure18}(b) and ~\ref{fig:figure12}.
\FloatBarrier

\section{Conclusions}\label{conclusions}
The main objective of this paper is to understand the homo-interactions between two droplets that are falling and two bubbles that are rising in a strongly coupled plasma medium. This medium has been considered as a visco-elastic fluid using the generalized hydrodynamic fluid model formalism. This work (part III) is a continuation of our earlier published work, parts I and II. In Part I~\citep{dharodi2021numericalA}, we separately explored the dynamics of a rising bubble and a falling droplet, and in Part II \citep{dharodi2021numericalB}, we explored the hetero- (bubble-droplet) interactions between a rising bubble and a falling droplet. In order to understand the home interactions, a series of two-dimensional numerical simulations have been carried out. Three different spacings between two droplets are simulated: widely, medium, and closely. In each case, the coupling strength has been presented as mild-strong ($\eta$=2.5, $\tau_m$=20) and strong ($\eta$=2.5, $\tau_m$=10). Since we observe that the behavior of two falling droplets is the similar as the behavior of two rising bubbles so will discuss here mainly droplet case. A few significant findings are as follows:
\begin{enumerate}
\item {Widely spaced}
\subitem{-~ In case of dusty plasma, unlike the HD fluid, emerging shear waves facilitate the pairing between two falling droplets. Due to the emerging shear waves the inner lobes get pinched in between the outer lobes. The  closeness of the inner lobes produces a new dipolar structure that is characterized by vertical upward motion. The outer lobes propagate outward, which is perpendicular to the original propagation direction. In plasma with mild strong coupling strength the  vertical upward motion of inner dipole remain continued. but for a plasma with strong coupling strength, the quicker-emerging waves removing energy from the lobes at a faster rate that reduced  the rotating influence of the lobes is more quickly. Consequently, the upward-moving inner blob pair begins to fill the gravitational supremacy and suddenly starts moving in a downward direction.}
\item {Medium spaced }
\subitem{ -~ Here, the coupling between the inner lobes results in two new dipolar structures which have unequal-strength lobes. These new two dipoles exhibits circular motion instead of vertical fall, where they collide with each other and exchange partners. After this, the outer lobes fall down along the gravity, while the inner lobes begin to move upward against gravity. The mild strong coupling strength medium, similar to the HD scenario, we have observed that two new dipoles of unequal strength blobs exhibit a circular motion and  exchange the blobs. But here both the outer and inner blobs start moving downward along the direction of gravity. The strong coupling strength medium, not like mild example, the inner blobs deform into filament shapes around the outer blobs.}
\item {Closely spaced}
\subitem{ -~The unlike-sign inner lobes of vorticities almost overlap with each other. So, under gravity, the pair of two droplets shows the downward dynamics like a single pair. In mild coupled plasma, compared to the HD case, the wake-type structure is diminished, the lobes are better separated, and the vertical upward motion gets reduced. In strong coupled plasma, the faster shear waves result in slower lobe vertical propagation and greater horizontal spacing between the blobs.}
\end{enumerate}
In summery, the net dynamic is governed by the competition between the mutual attraction of two inner like-sign vorticity lobes and the forward vertical motion of dipolar vorticities (unlike-sign lobes) due to gravity. In this paper, we confine our study of bubble-bubble/droplet-droplet interactions to the homogeneous background density medium. However, it would be exciting to extend this to  heterogeneous background density medium. The dynamics in a heterogeneous medium will get more complex due to the varying speed of shear waves with density \citep{dharodi2020rotating}. Furthermore, adding compressibility to the existing model will make it more like a real system. It would be interesting to observe how the dynamics of the rising bubble or falling-droplet are affected by the energy exchange between both modes and by newly developing waves.
 
\section{Acknowledgements}
This work was supported by the National Science Foundation through NSF-2308742 and NSF EPSCoR OIA-2148653.
\label{ack}
%


\end{document}